\documentclass[final, 5p, twocolumn, sort&compress]{elsarticle}

\usepackage{amssymb}
\usepackage{amsmath}
\usepackage[switch]{lineno}
\usepackage{units}
\usepackage{isotope}

\usepackage[pdftex, colorlinks, linkcolor=black, citecolor=black, urlcolor=black, breaklinks=true, hypertexnames=true, pdfauthor=B.~Zatschler]{hyperref}

\usepackage{microtype} 

\usepackage{caption} 
\usepackage{booktabs} 
\usepackage{multirow}

\journal{Nuclear Instruments and Methods in Physics Research Section A}

\begin{document}

\begin{frontmatter}

\title{Applying \textsc{Geant4}'s Importance Biasing to improve the efficiency of SuperCDMS background simulations}

\author[UofT,LU,SNOLAB]{B.~Zatschler\corref{correspondingauthor}}
\cortext[correspondingauthor]{Corresponding author}
\ead{birgit.zatschler@snolab.ca}

\author[TAMU]{A.J.~Biffl}
\author[SMU]{R.~Calkins}
\author[UofT]{M.D.~Diamond}
\author[LU,SNOLAB]{J.~Hall}
\author[UofT]{S.A.S.~Harms}
\author[TAMU]{M.H.~Kelsey}
\author[UdeM]{D.S.~Pedreros}
\author[UofT,LU,SNOLAB]{S.~Zatschler}

\affiliation[UofT]{organization={Department of Physics, University of Toronto},
             city={Toronto},
             postcode={M5S 1A7},
             state={ON},
             country={Canada}}

\affiliation[LU]{organization={Laurentian University},
             addressline={935 Ramsey Lake Rd},
             city={Sudbury},
             postcode={P3E 2C6},
             state={ON},
             country={Canada}}
             
\affiliation[SNOLAB]{organization={SNOLAB, Creighton Mine \#9},
             addressline={1039 Regional Road 24},
             city={Sudbury},
             postcode={P3Y 1N2},
             state={ON},
             country={Canada}} 
            
\affiliation[UdeM]{organization={Departement de Physique,  Universite de Montreal},
             city={Montreal},
             postcode={H3C 3J7},
             state={ON},
             country={Canada}}      

\affiliation[SMU]{organization={Department of Physics, Southern  Methodist University},
             city={Dallas},
             postcode={75275},
             state={TX},
             country={USA}}  
             
\affiliation[TAMU]{organization={Department  of  Physics  and  Astronomy,  and  the  Mitchell  Institute  for  Fundamental  Physics  and  Astronomy, Texas  A\&M  University,  College  Station},
             postcode={77843},
             state={TX},
             country={USA}}

\begin{abstract}

Experiments searching for extremely rare events surround their sensitive detectors with several layers of different shielding materials to protect them from external radiation and to achieve their low-background requirements to be able to observe a potential signal.
Standard Monte Carlo simulations that propagate particles through the thick shielding, usually do not penetrate the shield in sufficient numbers to properly model the external background, which is crucial for understanding the experiment's background composition. 

\textsc{Geant4} is a widely used toolkit to simulate the passage of particles through matter and it offers various biasing techniques, among them being importance biasing, which has been intensively explored for application in background simulations for the SuperCDMS experiment. In this article, the basic working principle of importance biasing is explained. Furthermore, we provide guidance for developers for their own implementation of a biasing scheme. A new track property, the \textit{biasing index}, is introduced to allow different track topologies to be distinguished. Validation studies and optimal parameters for biasing gammas and neutrons are presented and caveats are discussed. In this work, simulations run with importance biasing achieved an efficiency boost of about $\mathcal{O}(10^4)$ for gammas and up to 500 for neutrons. 
By applying these techniques, we show that energy distributions simulated with and without importance biasing are consistent with each other within statistical uncertainty at a fraction of the consumed computing time.

\end{abstract}


\end{frontmatter}

\section{Introduction}

Low-background experiments often feature an extensive shielding made of $\sim \mathcal{O}(10~\unit{cm})$ of lead to absorb external gamma rays, and hydrogenous materials to stop neutrons from entering the sensitive detectors.
While these shieldings are designed to be very effective, they impose efficiency issues onto corresponding Monte Carlo simulations, because even for a huge number of simulated primary particles $\sim \mathcal{O}(10^{12})$, the number of particles reaching the detector are insufficient to estimate their rates. 
As a consequence, simulations of background sources within the experiment's materials, and in particular simulations of external radiation, lack sufficient statistics despite consuming a large amount of CPU time $\sim \mathcal{O}(100~\unit{CPU~years})$. This leads to large statistical uncertainties in simulated spectra, impeding background estimates for design studies or the construction of a background model.

To overcome these issues, \textsc{Geant4} \cite{Geant4-2016, Geant4-2006, Geant4-2003} offers various biasing mechanisms which can significantly enhance the simulation efficiency.
In particular, importance biasing can increase the number of particles reaching the sensitive detector through a rethrowing and reweighing scheme, which is described later in the paper.
This technique can improve the efficiency by several orders of magnitude for the same number of simulated primary particles.

\section{SuperCDMS}

\begin{figure}[t]
\centering
\includegraphics[width=\linewidth]{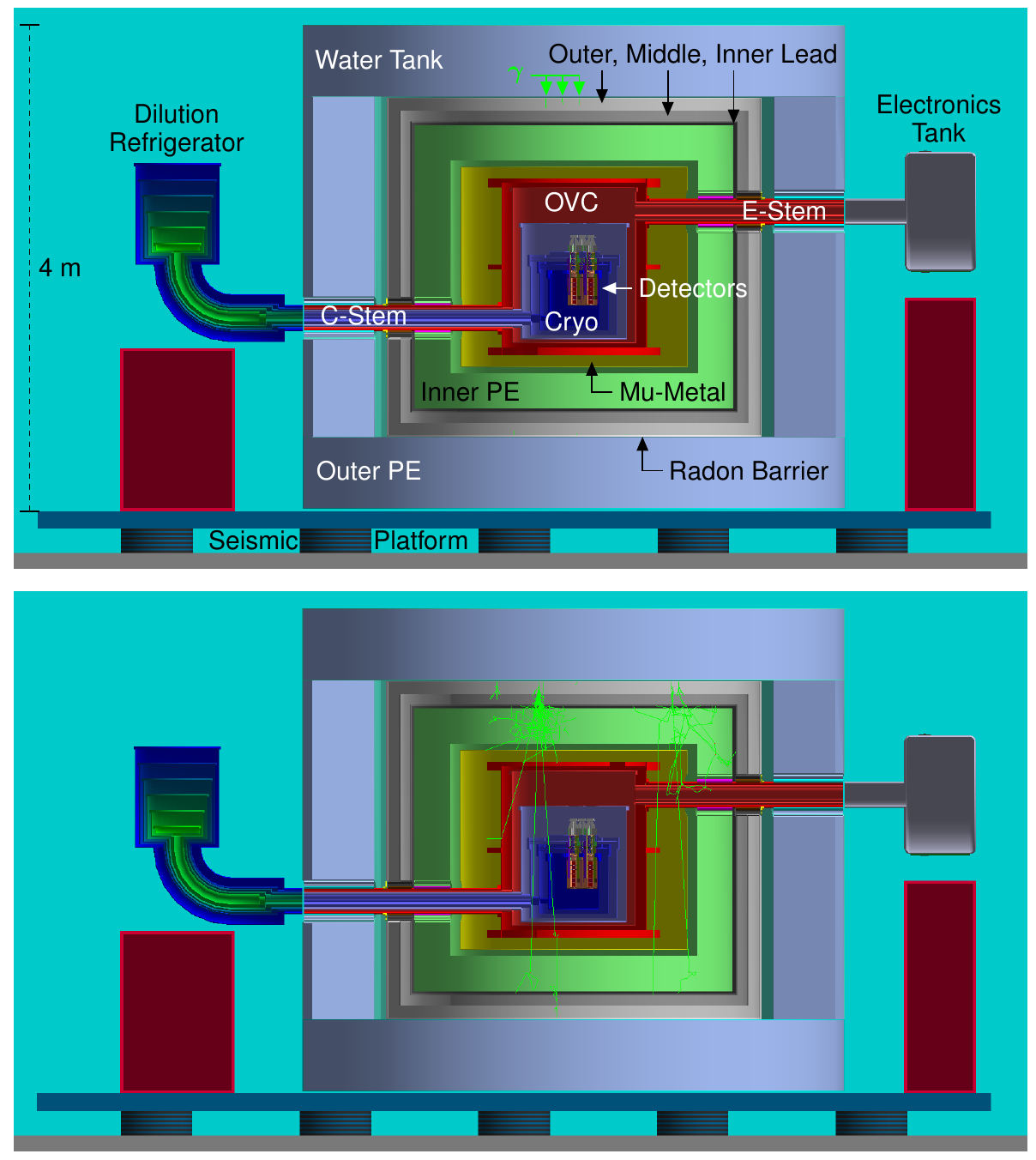}
\caption{
Visualization of ten gammas (green tracks, starting position indicated with $\gamma$), each having an energy of 2 MeV, started at the surface of the radon barrier of SuperCDMS with a momentum direction pointing inward to the cryogenic detectors at the center of the outer vacuum chamber (OVC). The particle propagation is shown without importance biasing (top) and with importance biasing (bottom) using 16 importance layers each having a thickness of 1.25~cm, thus spanning the full 20 cm of the lead shield.
In the normal simulations (top) all ten gammas are absorbed inside the lead shield. But with importance biasing (bottom) some events generate a visual shower of gammas inside the lead shield resulting in many particles which fully traverse the thick shield volume.
}
\label{fig_supersim_supercdms}
\end{figure}

The Super Cryogenic Dark Matter Search (SuperCDMS) experiment is a direct detection dark matter (DM) experiment with the design goal of a sensitivity to DM-nucleus scattering within a mass range of 0.5 to $5~\unit{GeV/c^2}$. SuperCDMS will operate a total of 24 cryogenic Ge and Si crystals 2~km underground at SNOLAB, located in Sudbury, Canada. That deep underground, the muon flux is reduced by a factor of 50 million compared to Earth's surface \cite{SNO-2009}. Additionally, the SuperCDMS detectors are surrounded by an inner polyethylene (PE) layer of 30~cm, 20~cm of lead, a 5~mm radon barrier of aluminum as well as 60~cm of PE (bottom) and water (top and sides) to shield them from external gamma rays and neutrons originating from the rock cavern walls. Fig.~\ref{fig_supersim_supercdms} depicts a visualization of the main components implemented in SuperCDMS' \textsc{Geant4} application \textit{SuperSim}.

\section{Importance Biasing}

Importance biasing is a common variance reduction technique which is utilized in Monte Carlo particle transport codes, including, but not limited to \textsc{Geant4}, MCNP \cite{MCNP-Man} and FLUKA \cite{FLUKA}.

\textsc{Geant4} provides various event biasing techniques: geometrical methods; physics based which adapts cross sections or affects desired processes; Reverse Monte Carlo where particles are generated in the sensitive detectors and are tracked backwards; and also a generic biasing method allowing the user to create a dedicated, individual biasing method \cite{Geant4-AppDev}.

In this article, we focus on the geometrical method called importance biasing. In our implementation in SuperSim, we make use of \textit{modular physics lists}, which is based on the example \textit{biasing/B02} provided with the \textsc{Geant4} software package \cite{Geant4-AppDev}.

SuperCDMS' implementation has been originally developed for gammas, however, it can also be applied to neutrons with adapted settings which will be pointed out in the respective sections. 
A conference proceeding article describing only the basic implementation for gammas has been previously published in \cite{TAUP-Proc}.

\subsection{Basic working principle}

\begin{figure*}[t]
\centering
\includegraphics[width=14cm]{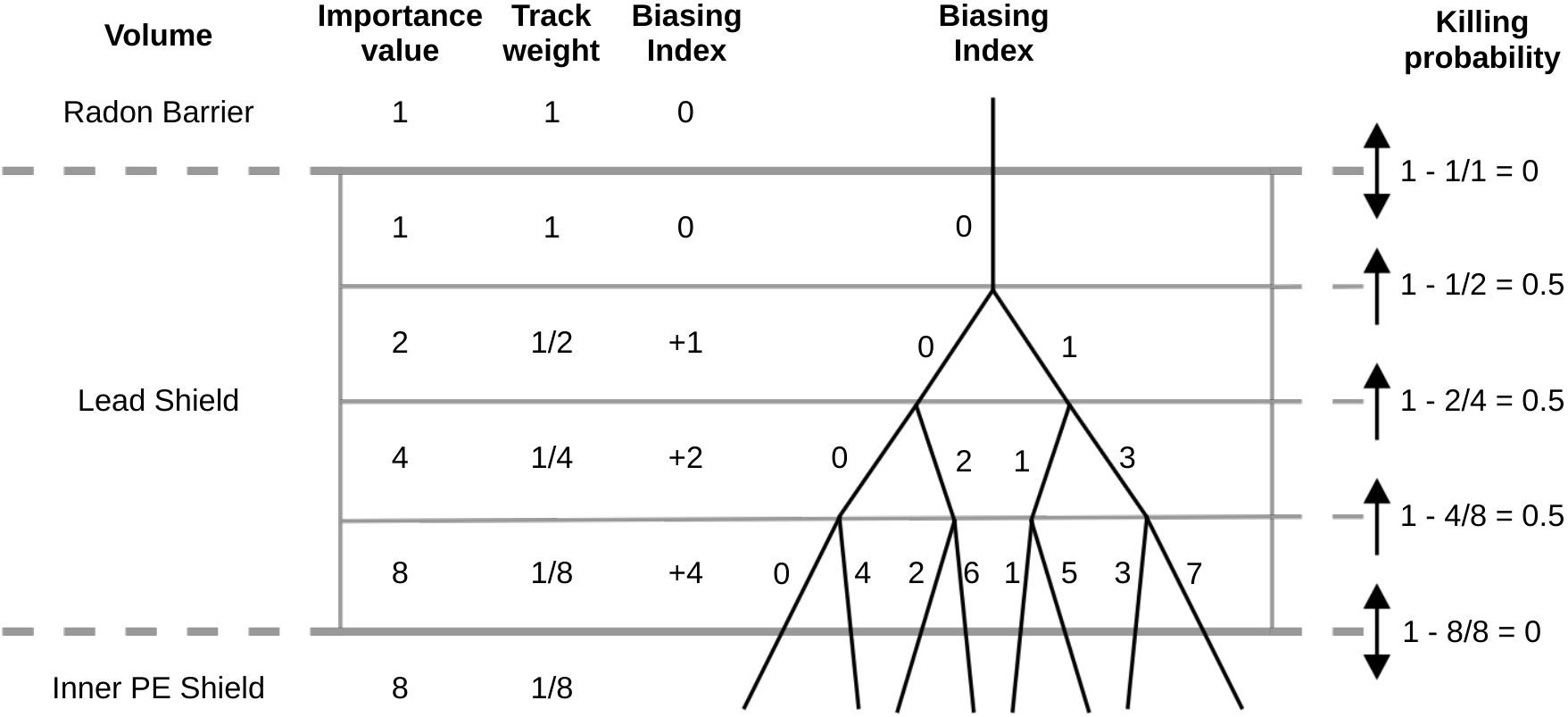}
\caption{Overview of the basic working principle of importance biasing with an example of an importance splitting factor of two using four importance layers overlapping with the SuperCDMS lead shield. The biasing index distinguishes the different track topologies.}
\label{fig_imp_bias_working_principle_biasing_index}
\end{figure*}

Fig.~\ref{fig_imp_bias_working_principle_biasing_index} shows a schematic of the working principle of importance biasing. It introduces \textit{importance layers}, which are virtual geometries in a parallel world overlaid on the SuperCDMS lead shield geometry in the real (mass) world. Each importance layer $L$ has been assigned an importance value $V$, increasing from outside to inside with $V = 2^L$, starting with $L=0$.

Each time a biased particle crosses the boundary between two importance layers, for which the importance value ratio between the next and the previous layer is~2, the particle is duplicated, i.e., copied. The copied particle has the exact same properties as the original particle, i.e., the same energy, momentum direction, position, etc. However, the track weight of both particles is divided by~2, i.e.\ the total track weight is conserved.
Both particles are tracked independently of each other and continue with different random engine states, hence their next steps are not identical.

If a biased particle moves away from the sensitive detectors and it crosses the boundary between two importance layers where the ratio between the next and the previous layer is 0.5, there is a 50\% probability that the particle track is killed. In the case the particle survives, its track weight is multiplied by~2.

The pictured example uses an importance splitting factor of two, but any other factor could be used as long as the importance values are greater than zero. E.g.\ with an importance splitting factor of four, the importance value ratio between the next and the previous value would be four, multiplying the biased particle by four. In the other direction, the importance value ratio would be 0.25, i.e.\ the probability that the biased particle is killed is 75\%. Statistically, different importance splitting factors will result in the same answer, but the efficiency of the simulations could vary.
The choice of the importance splitting factor also affects the optimal settings for the other importance biasing parameters, which again affects the efficiency (see section~\ref{sec_optimal_settings}). 

The described biasing techniques, i.e., duplicating and killing particles at importance layer boundaries, lead to more particles being tracked which are traveling into the direction of the sensitive detectors, eventually increasing the number of detector hits, and fewer particles being tracked which are moving away from the detectors, saving computing time by not simulating particles which have a geometrically low probability to reach the sensitive detectors.

Fig.~\ref{fig_imp_bias_working_principle_biasing_index} also indicates that for the example of $L=4$ importance layers, there can be up to $2^{L-1} = 8$ different track topologies which need to be distinguished to create a valid detected spectrum. If the track topologies in an event are not distinguished, one could observe non-physical energy collection due to interactions from different kinds of track topologies.

\subsection{Biasing Index}
\label{sec_biasing_index}

We introduce a new track property, the biasing index $\mathcal{B}$ to distinguish between different track topologies and combine the detector hits of tracks belonging to the same topology.
Fig.~\ref{fig_imp_bias_working_principle_biasing_index} illustrates the algorithm used to calculate and propagate the biasing index.

Every primary particle starts with a biasing index of 0 and a track weight of 1, independent of its origin location. The same is true if there is more than one primary particle in a \textsc{Geant4} event (\texttt{G4Event}).
Each time a particle is copied, the original particle keeps its biasing index, while the copied particle gets assigned a new value. The biasing index $\mathcal{B}_\text{copy}$ of the copied particle is calculated by taking into account the original particle's biasing index $\mathcal{B}_\text{original}$ and the importance value $V$ of the importance layer into which the original particle has just moved:

\begin{equation}
\label{eq_biasing_index}
\mathcal{B}_\text{copy} = \mathcal{B}_\text{original} + V/2
\end{equation}

In all other physics processes which create secondary particles, the biasing index and track weight are inherited from the parent particle.

In the analysis of biased simulations, which is in our case not a part of the \textsc{Geant4} simulation itself, tracks having identical biasing indices in the same \texttt{G4Event} are combined to determine the total energy deposit of the event topology, while tracks with different biasing indices within the same \texttt{G4Event} are handled as separate events. Tracks generated from different primary particles within the same \texttt{G4Event} are combined if their biasing indices are identical. Note that tracks having the same biasing index also have the same track weight after crossing all importance layers (see Fig.~\ref{fig_imp_bias_working_principle_biasing_index}) which is essential for a correct analysis.
To reconstruct a spectrum of the observed energy deposits in the sensitive detectors, each particle track is weighted with its track weight, i.e.\ in a histogram the corresponding total energy deposit of the event enters with the count weight being the track weight.

All the above also applies to primary particles which are started within an importance layer. In such cases, a particle starts with a track weight of~1 in an importance layer with an importance value of e.g.~2 and the pictured example tracks in Fig.~\ref{fig_imp_bias_working_principle_biasing_index} shift down by one importance layer, ending with only up to four different biasing indices. In the case of more than one primary particle in a \texttt{G4Event}, it is important to note that the resulting track weights and biasing indices are only consistent if all primary particles started in the same importance layer.

\subsection{Special cases and biasing index for backwards going tracks}
\label{sec_special_cases}

Since only one particle type is biased in our case, which are in the following examples gammas, other particle types are not affected by the importance basing. This means that if e.g.\ an electron crosses the importance layer boundary, it gets neither split nor killed, nor is its track weight manipulated. 
This also means that the electron takes its current biasing index unaltered over the boundary which leads in the end to tracks having the same biasing index, but different track weights.
Fortunately, these cases are rather rare, so we just discard the affected tracks at the analysis level. Due to the differing track weight, such cases are easily detectable.

\begin{figure*}[t]
\centering
\includegraphics[width=14cm]{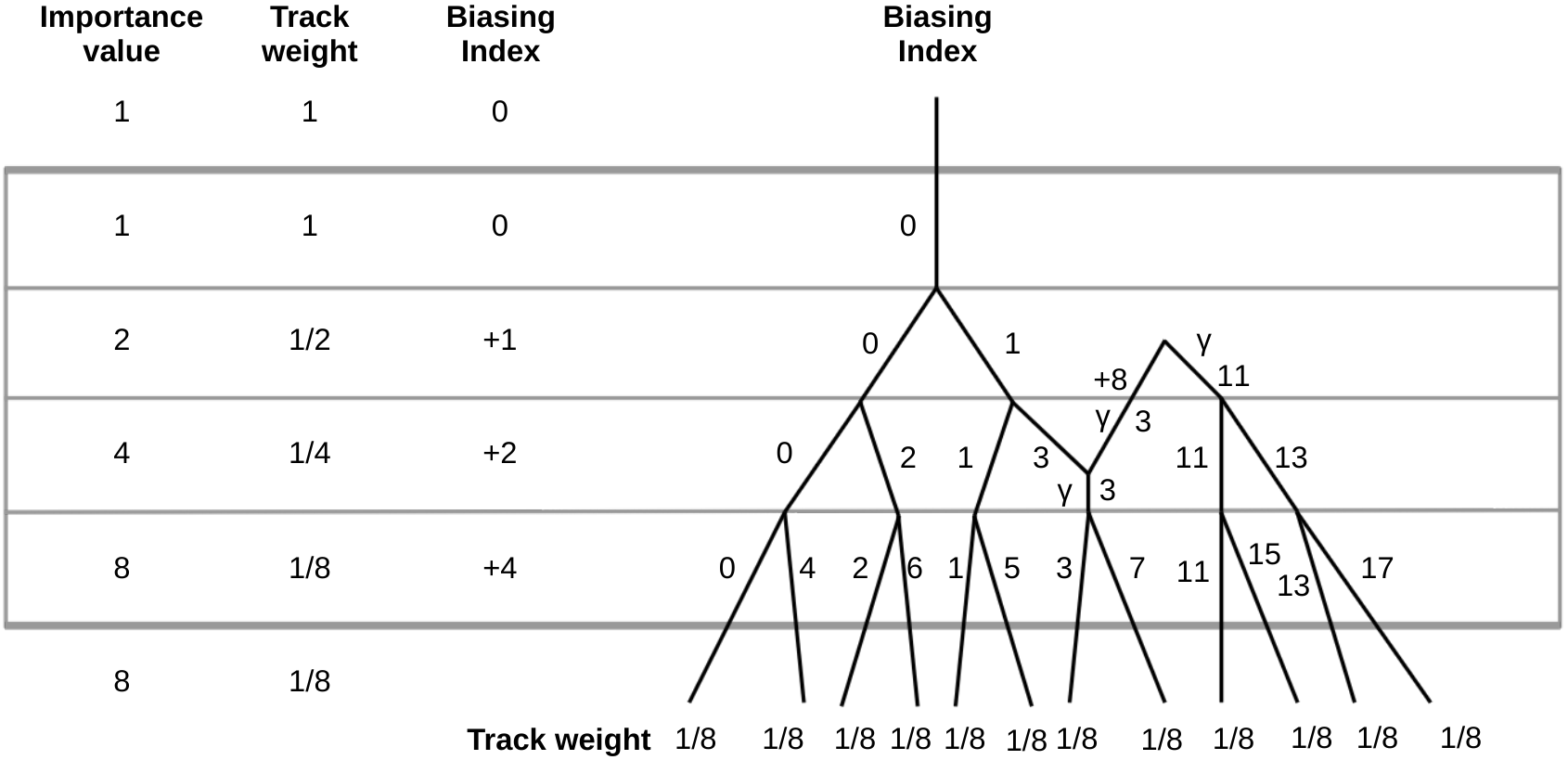}
\caption{Example for a biased particle ($\gamma$) going backwards through an importance layer boundary and getting assigned a new  biasing index by adding the maximum importance value. Without this, the track with $\mathcal{B}=13$ would have a biasing index of 5 even though it does not belong to the topology of the other track with $\mathcal{B}=5$.}
\label{fig_imp_bias_backwards_biasing_index}
\end{figure*}

Another case that can happen is a gamma crossing an importance layer boundary backwards, i.e.\ the ratio of the importance value of the next layer divided by the previous layer is 0.5. If the gamma survives, gets scattered and crosses the boundary again, but this time forwards, it can mess up the biasing index in such a way that tracks have the same biasing index but  do not belong to the same topology. 

To account for these cases, the backwards biasing index is introduced. Each time a biased particle goes backwards through the boundary (and if it survives), its biasing index is increased by the maximum importance value. Fig.~\ref{fig_imp_bias_backwards_biasing_index} depicts an example, with the maximum importance value being~8. With that, tracks with different topologies do not end up with the same biasing index.

There are caveats though, as it is possible that tracks of the same topology do not get combined, because they have different biasing indices. However, this is extremely rare for gammas due to their physics processes. 
One gamma would have to go backwards, while another one of the same topology goes forwards.

For just one primary gamma in the investigated energy range, there are two typical processes to create two gammas within the same topology. If the energy of the gamma is above 1.022~MeV, i.e.\ two times the electron mass, it can undergo pair production and the resulting positron annihilates with an electron, creating two gammas of 511~keV. Another possibility is that a gamma  is Compton scattered, transferring some of its energy to an electron, which can in turn create Bremsstrahlung.
Both of these processes create rather low-energy gammas and it is highly improbable that both reach the sensitive detectors.

Even in the case of two high-energy gammas within the same topology, e.g.\ emitted by \isotope[60]{Co} with a certain angular correlation, at least one of them must undergo Compton scattering with a large change in momentum direction to travel into the same direction as the other gamma. In that process, the scattered gamma would transfer a good part of its kinetic energy, effectively reducing its chances to reach a sensitive detector thereafter.

However, if neutrons are biased, these effects occur more frequently and become apparent in the validation spectra as large residuals between biased and unbiased simulations (see Fig.~\ref{fig_validation_neutrons} and \ref{fig_optimal_settings_neutrons} in section~\ref{sec_validation}). A single neutron can undergo the $(n,2n)$ process, multiplying the number of neutrons. Since neutrons scatter much more often and have a much longer range compared to gammas, it does happen more regularly that tracks of the same topology end up with different biasing indices and thus the total deposited energy of this topology is divided among them.

The frequency of the described special cases is strongly dependent on the importance biasing settings, in particular the thickness of the importance layers and their quantity. Section \ref{sec_optimal_settings} describes how to find optimal settings to avoid a number of undesired effects and achieve a sufficient improvement of the simulation efficiency.

\subsection{Technical notes and implementation}

This section serves as a guide for \textsc{Geant4} application developers who want to implement importance biasing for their own simulations. Since applications can be set up differently according to the user's needs, not all of the following \textsc{Geant4} class references might be accurate or apply to every application.

The example \textit{biasing/B02}, released with \textsc{Geant4} version 10.7.4, has been used as the basis for SuperCDMS' implementation in SuperSim. Some code snippets from SuperSim can be also found in Ref.~\cite{VIEWS-Talk}.
An initial implementation attempt based on example \textit{biasing/B03} is described in Ref.~\cite{David-Thesis}, but it has been replaced later on because it only worked in single-threaded mode, while we run SuperSim in multi-theaded mode.
For SuperCDMS' simulations, we use the \texttt{Shielding} physics list with \texttt{G4EmStandardPhysics\_option4} and the \texttt{NeutronHP} model for low energy electromagnetic and neutron interactions, respectively \cite{Geant4-Phys}.

\paragraph{\textbf{Parallel world}}
A parallel world is necessary for placing the importance layers.  \textsc{Geant4} can handle more than one parallel world and importance biasing should have its own parallel world. We name it \texttt{"ImportanceBiasing"} in the following. If more than one particle type shall be biased at the same time, each particle type needs its own parallel world.

When setting up the \texttt{G4GeometrySampler}, the particle type, which shall be biased, needs to be selected and the corresponding parallel world has to be chosen. If more flexibility is desired, the particle type selection can also be done via a macro command.

\paragraph{\textbf{Physics lists}}
The implementation in the example utilizes modular physics lists (\texttt{G4VModularPhysicsList}). If the developer's \textsc{Geant4} application also already uses modular physics lists, the corresponding \texttt{G4ImportanceBiasing} and \texttt{G4ParallelWorldPhysics} can be directly added to the existing physics lists.
In any case, it is recommended to toggle importance biasing via a macro command in the physics list messenger as only certain simulations will benefit from it.

\paragraph{\textbf{Importance layers}}
In the application's parallel world class, which inherits from \texttt{G4VUserParallelWorld}, the importance layers need to be constructed just like every other solid, logical and physical volume, however, no material needs to be assigned as in the mass world (see also section ``Parallel Geometries'' in Ref.~\cite{Geant4-AppDev}).

An importance store (\texttt{G4IStore}) is created for the parallel world instance \texttt{"ImportanceBiasing"}. The physical volume of the parallel world gets assigned an importance value $V=1$. From outside to inside, each importance layer $L$ gets $V=2^L$, starting with either $L=0$ or $L=1$. The examples shown in Fig.~\ref{fig_imp_bias_working_principle_biasing_index} and \ref{fig_imp_bias_backwards_biasing_index} start with $L=0$. The physical volume of the innermost importance layer needs to completely cover the full interior space without any holes, so that the importance value inside is the largest value as indicated in the schemata.

\paragraph{\textbf{Tracking action}}
The developer's application needs a class which inherits from \texttt{G4VUserTrackInformation} to introduce the biasing index as a new track property, which should be propagated to the application's hit collection to be stored in the simulation's output.
The calculation of the biasing index can be done in the developer's class inheriting from \texttt{G4UserTrackingAction}, preferably in the \texttt{PostUserTrackingAction}. There, the secondary particles of the track in question can be retrieved. If the creator process name of a secondary particle is \texttt{"ImportanceProcess"} (name fixed by \textsc{Geant4}), then Equation~\ref{eq_biasing_index} can be applied and the new biasing index can be assigned to the secondary particle in its track information.

\paragraph{\textbf{Stepping action}}
The biasing index for backwards going particles must be assigned dynamically. A good place for that is inside the application's stepping action inheriting from \texttt{G4UserSteppingAction}. There, the developer has access to each step and can check whether a step was limited by the importance biasing process and if the particle was crossing an importance layer boundary backwards. If this applies, the particle's track information can be retrieved and the biasing index can be increased by the maximum importance value.
Note that it might not be intuitive that the name of the process limiting the step is the same as for the parallel world, i.e.\ \texttt{"ImportanceBiasing"} in our case, which is different from the above mentioned \texttt{"ImportanceProcess"} for the creator process name of the secondary particle.

If the biasing index for backwards going particles is set as described, this must be reflected in the application's tracking action class where the calculation of the biasing index for the secondary particles is happening. This is because the biasing index of a backwards going track is not consistent for the whole track (see also Fig.~\ref{fig_imp_bias_backwards_biasing_index}). Instead, the biasing index $\mathcal{B}_\text{original}$ of the original track in Equation~\ref{eq_biasing_index} refers to the particle's biasing index at the step where the secondary particle was created.
Thus, it is useful to store whether a track went backwards, and in particular how many secondary particles have been created in each step of this track, in the application's track information class. With that information it is possible to calculate the biasing index that a track had at each step in the application's tracking action.

\paragraph{\textbf{Track weight}}
Finally, the developer does not need to take care of the particle's track weight as this is calculated automatically by \textsc{Geant4} every time the importance biasing process limits the step, i.e.\ when a particle crosses an importance layer boundary.

\section{Validation studies}
\label{sec_validation}

For the validation that the importance biasing and biasing index implementation work as intended, simulations were run with and without importance biasing for each isotope, decay chain and neutron spectrum which are supposed to be simulated to model SuperCDMS' background. Note that we are simulating neutrons separately from their original decay chains by drawing from spectra generated by SOURCES4C \cite{SOURCES4C}. To avoid double counting, we turn off the generation of neutrons by e.g.\ spontaneous fission or the $(\alpha,n)$ process in the decay chain simulations with \textsc{Geant4}.

The validations have been performed for single and multiple primaries within the same \texttt{G4Event} because many typical background sources emit more than just one particle, e.g.\ \isotope[60]{Co} emits two coincident gammas of similar energies and intensities and \isotope[208]{Tl}, which is part of the decay chain of \isotope[232]{Th}, has a rather complex gamma cascade consisting of various gammas energies and intensities. Additionally, the natural decay chains can emit a number of neutrons via spontaneous fission and more neutrons may be produced in ($\alpha, n$) reactions.

In the first iteration of validation simulations, solely gammas and neutrons are started as primary particles and are biased, respectively. In section \ref{sec:eff_boost}, the results for isotopes, decay chains and neutrons from decay chains are discussed.
Since the unbiased simulations do not deliver sufficient statistics when comparing detector hits, several adaptions have been made for the validation studies.

When biasing gammas, the primary particles are started at the radon barrier (see Fig.~\ref{fig_imp_bias_working_principle_biasing_index}) and are propagated through the importance layers which cover the lead shield. After the innermost importance layer, i.e., at the outside of the inner PE shield, all particle tracks, their energy and track weight are recorded and the resulting spectra are constructed taking into account the biasing index.
Note that not all particles which reach the innermost importance layer at the inner PE shield will also reach the detectors.

\begin{figure}[t!]
\centering
\includegraphics[width=\linewidth]{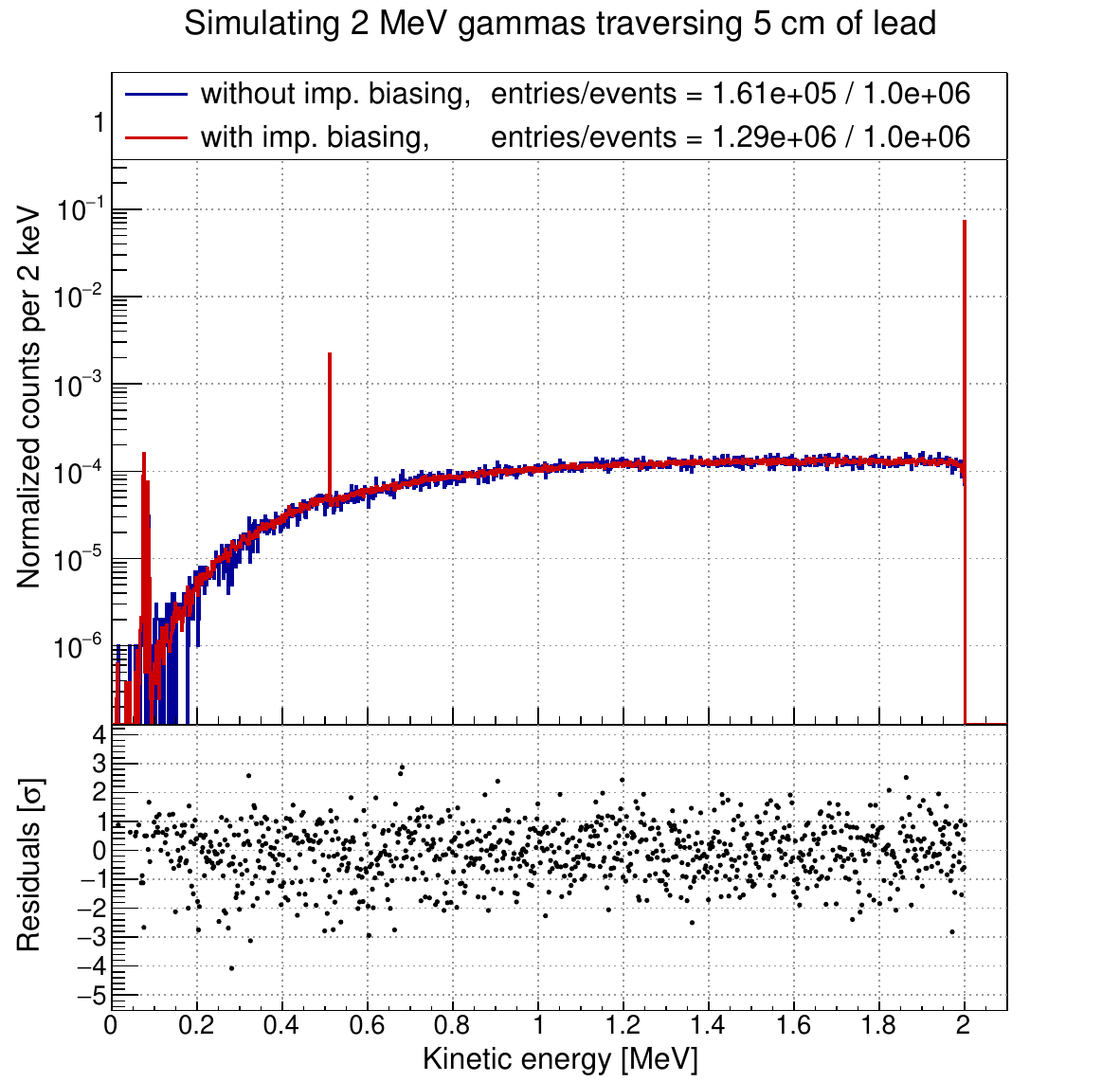}\\
\includegraphics[width=\linewidth]{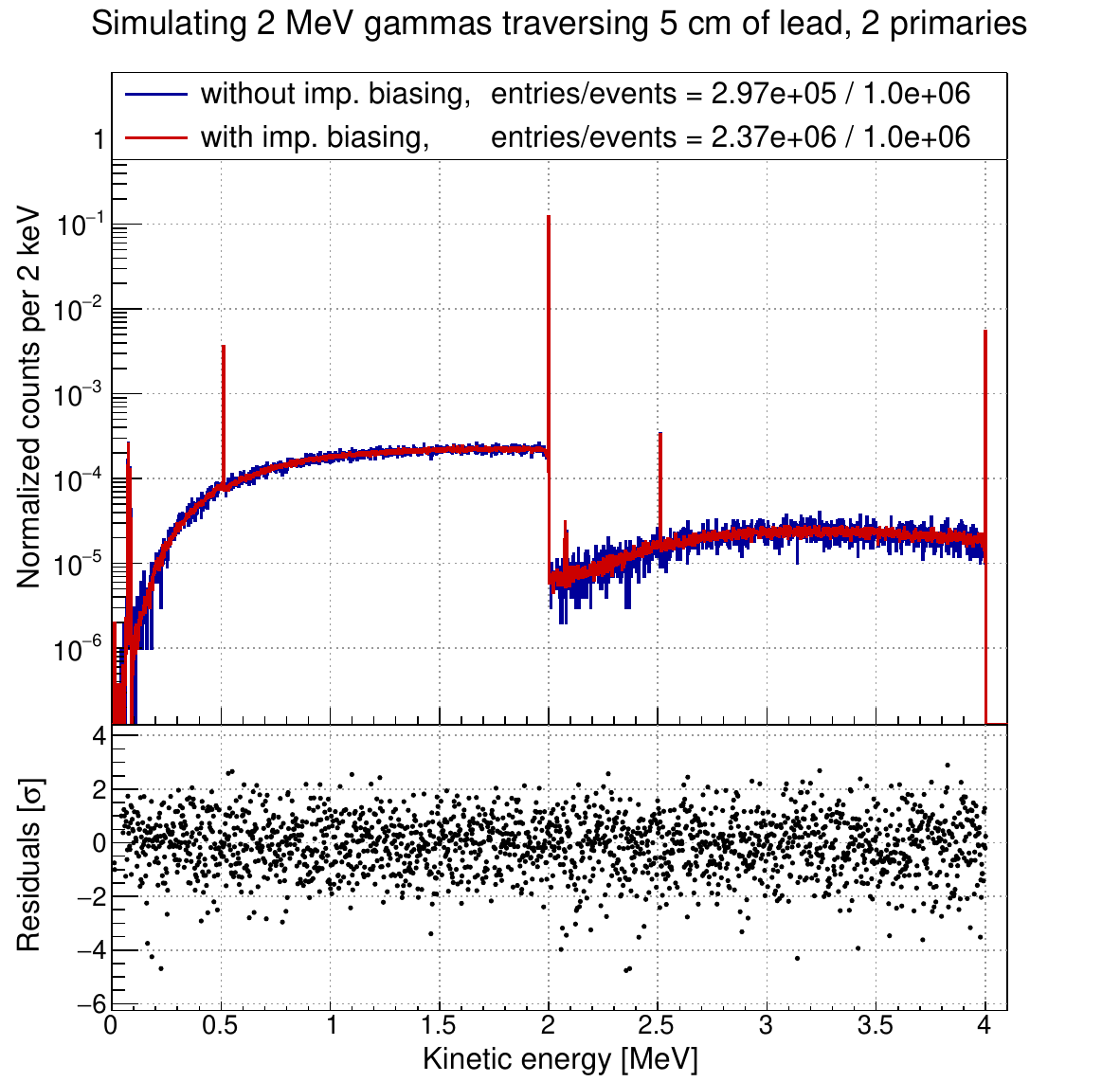}
\caption{
Comparison of unbiased and biased simulations for 2 MeV gammas going through 5~cm of lead for one primary (top) and two primaries (bottom) per event. The legend shows the number of entries and the number of simulated events for each simulation. The biased simulations were run with four importance layers, each having a thickness of 1.25~cm. The residuals show that all features are well modeled in the biased simulation: X-ray peaks from lead around 80~keV, gammas at 511~keV from $e^+e^-$ annihilation, the full energy peak at 2~MeV, and also sum peaks of these features when simulating two primaries per event.
}
\label{fig_validation_gammas}
\end{figure}

Since the lead absorbs most gammas, it is not feasible to run simulations without importance biasing for comparison as the statistics will not be sufficient. Thus, the lead shield thickness is reduced from its original 20~cm to 5~cm for these validation studies.
Fig.~\ref{fig_validation_gammas} shows the comparison of biased and unbiased simulations for 2 MeV gammas. The counts in each histogram are normalized to the number of simulated events. In the biased simulations, the counts are already weighted by the track weight. Residuals $R_i$ are calculated with the normalized counts $C$ per bin $i$ in each histogram:

\begin{equation}
\label{eq_residuals}
R_i = \frac{C_{i,\text{unbiased}} - C_{i,\text{biased}}}{\Delta C_{i,\text{unbiased}}}
\end{equation}
with the uncertainty $\Delta C_i = \sqrt{C_i}$ for Poisson distributed counts. 
In the case $\Delta C_{i,\text{unbiased}}$ is zero because there is no entry in the unbiased histogram, no residual is calculated for this bin.
By definition, the residuals $R_i$ should have approximately a normal distribution with a mean of zero and a standard deviation $\sigma$ of one for equivalent Poisson distributions.

\begin{figure}[t!]
\centering
\includegraphics[width=\linewidth]{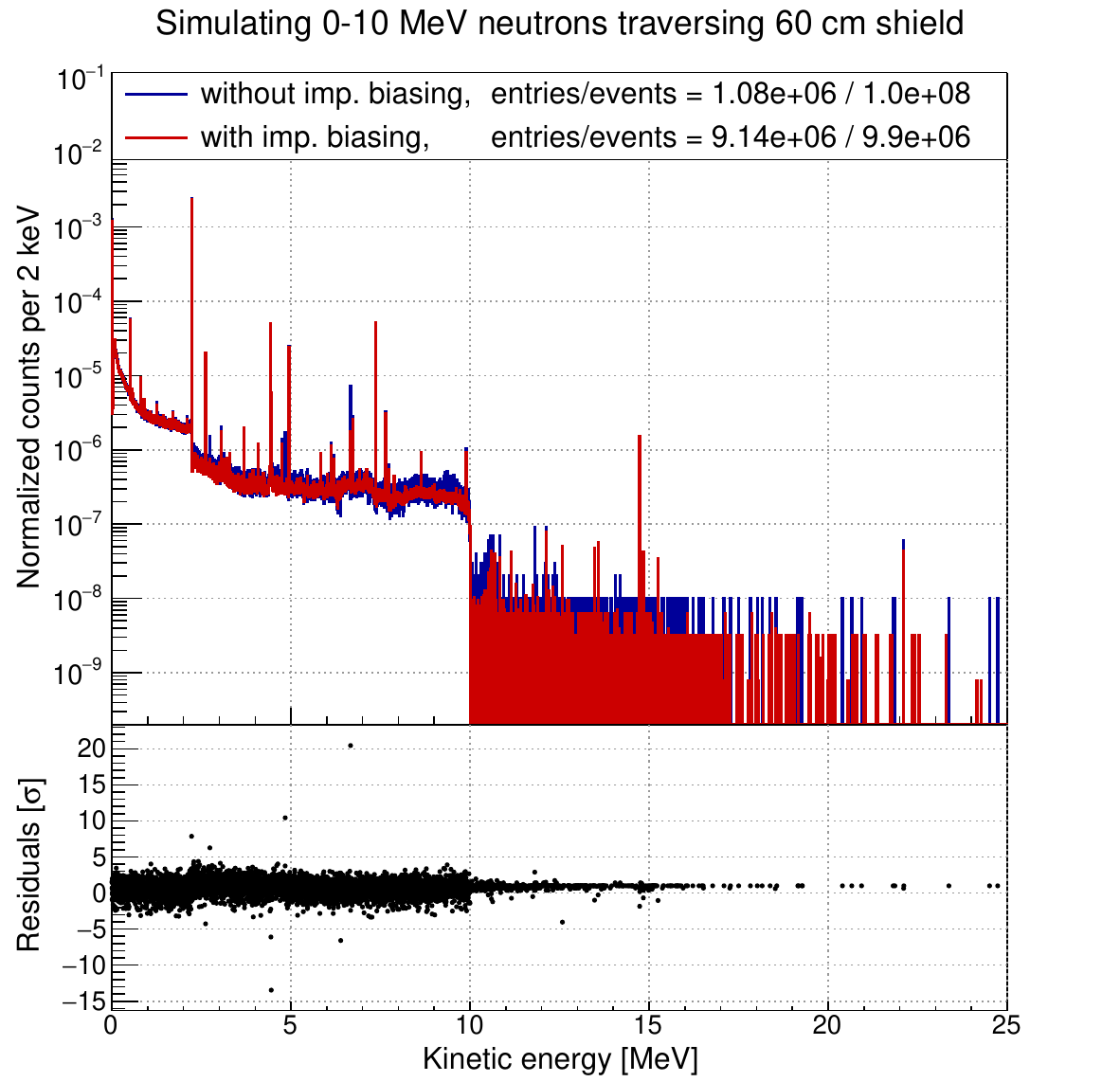}\\
\includegraphics[width=\linewidth]{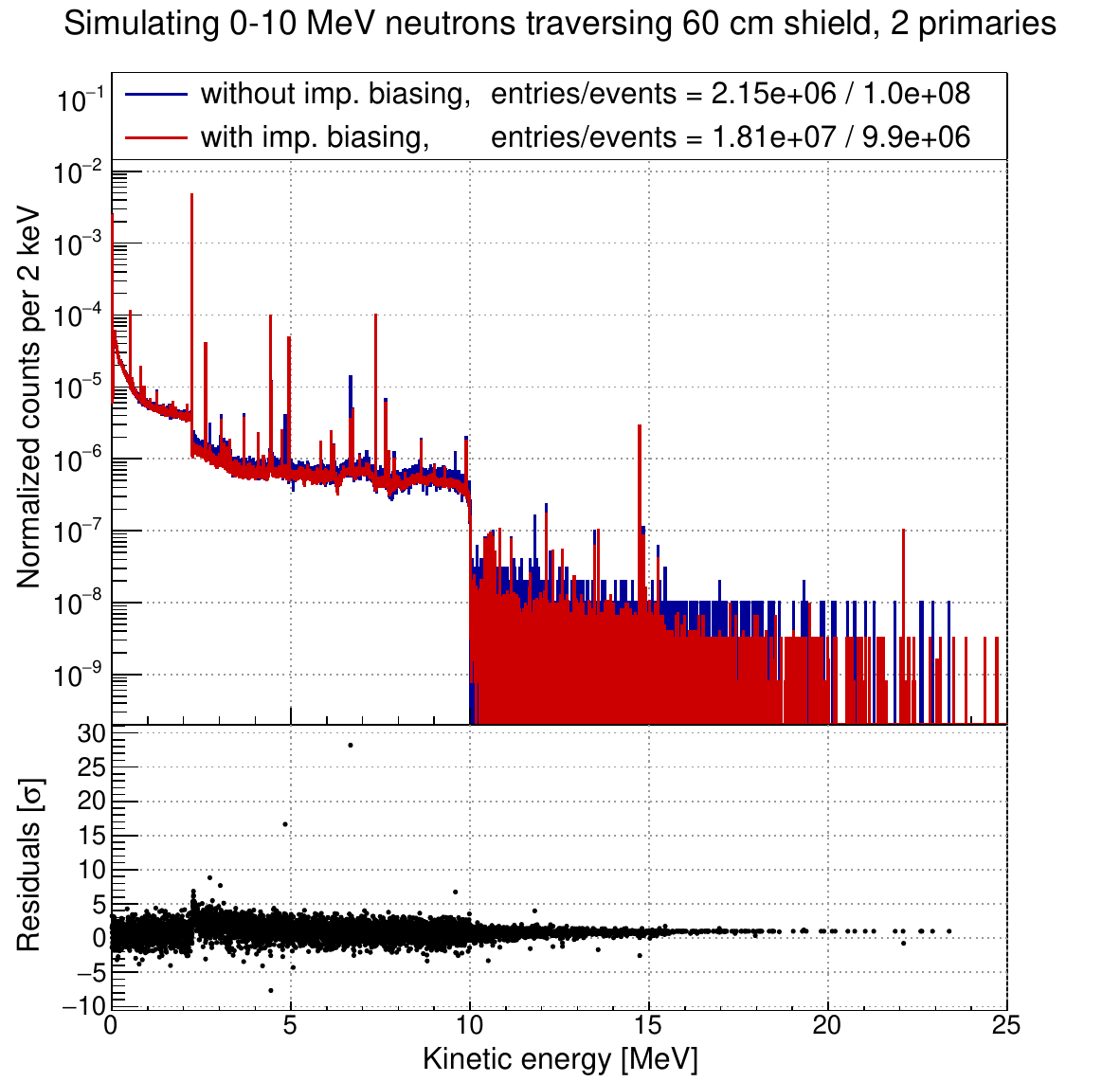}
\caption{
Comparison of unbiased and biased simulations for neutrons uniformly distributed between $0-10~\unit{MeV}$ for one primary (top) and two primaries (bottom) per event. The biased simulations were run with eight importance layers, each having a thickness of 7.5~cm. The peak features in the spectra are gammas with discrete energies produced in neutron captures $(n,\gamma)$ and neutron inelastic scattering $(n,n')$ in the different materials composed of H, C, O, Al, Fe, Pb, as well as gammas emitted during the decay of activated isotopes from these processes. Depending on the specific process, these gammas or gamma cascades can have higher energies than the initial neutron.
The residuals show a general match of biased and unbiased spectra, however, some peaks are less intense in the biased spectra. Such deviations are expected (see section~\ref{sec_special_cases}) and deemed to be acceptable for our purposes.
}
\label{fig_validation_neutrons}
\end{figure}

When biasing neutrons, the primary particles are sampled at the outside of the water tank and outer PE shield, while the importance layers span over the total shield thickness consisting of the water tank and outer PE shield, the radon barrier, lead shield and inner PE shield. Particle tracks are recorded at the outside of the mu-metal shield (see Fig.~\ref{fig_supersim_supercdms}).
The neutron validation employed a similar strategy as used for the gammas, namely reducing each shield component to half of its original thickness, i.e.\ the total shield thickness is decreased from 120~cm to 60~cm to achieve sufficient statistics in the unbiased simulations.
Fig.~\ref{fig_validation_neutrons} shows the comparison of biased and unbiased simulations for a continuous uniform spectrum of neutrons in the range of $0-10~\unit{MeV}$.

For gammas (Fig.~\ref{fig_validation_gammas}) the unbiased and biased spectra agree very well throughout the full energy range, which is also reflected by the residuals clustering close around zero. But for neutrons (Fig.~\ref{fig_validation_neutrons}), in particular with two primary particles, in the region between 5 and 10~MeV, the residuals do not average to zero.
The main cause for this is that neutrons belonging to the same topology end up with different biasing indices since neutrons scatter around more frequently and have a longer range (also see section~\ref{sec_special_cases}).

In all cases, the starting positions of primary particles are limited to be far away from the C-Stem and E-Stem (see Fig.~\ref{fig_supersim_supercdms}). The importance layers are constructed as simple nested cylinders in the parallel world and also cover the stems even though there is only vacuum in them.
Hence, particles traveling through the stems would be multiplied due to the importance biasing, but not be absorbed by any material. This would result in a very large number of tracks being produced which are caused by just a few primaries which had been sampled close to the stems.

In the reconstructed spectra, such events would show up as peaks at random energies if the primary gamma had been Compton scattered. These peaks are equivalent to what one would observe in a simulation run without importance biasing, where a few tracks made it through the stems, which would be seen as single entries in a histogram.

The origin of these enhanced artifacts of the biasing technique are well understood, but they are misleading when validating biased against unbiased spectra. However, since our sensitive detectors are not in line of sight with the stems, such features wash out and there is no negative impact for background simulations generating particles near the stems.

Since the occurrence of such features is very dependent on the experiment's geometry and the initial gamma energy, a case-by-case study is recommended. If the detected spectra are affected by these artifacts, one could run separate simulations: with importance biasing sampling primaries far away from the stems and without importance biasing close to the stems.
In any case, on average the total rate through the stems is correctly captured by simulations ran with importance biasing.

\section{Efficiency Boost}
\label{sec:eff_boost}

A simulation run with importance biasing can be significantly more efficient compared to an unbiased simulation, but this strongly depends on the importance layer thickness and the number of importance layers.
First, the number of different biasing indices $N_{\mathcal{B},i}$ recorded for each simulated event $i$  is used to calculate the average number of different biasing indices per event:
\begin{equation}
\overline{N_{\mathcal{B}}} = \frac{ \sum_i N_{\mathcal{B},i}}{N_\text{events}} 
\end{equation}
In the unbiased simulation, $N_{\mathcal{B},i}$ can only be 0 or 1, while in the biased simulation it can be much larger.

The biased simulation propagates up to several orders of magnitude more tracks, hence on a per-event basis it also runs much longer than the unbiased simulation.
Consequently, the efficiency boost should be normalized by the average CPU time per event, which is given by the runtime $t$ and the number of simulated events $N_\text{events}$:
\begin{equation}
\varepsilon = \frac{ \overline{N_{\mathcal{B}\text{, biased}}} } {\overline{N_{\mathcal{B}\text{, unbiased}}}} \cdot \frac{ t_\text{unbiased} } { t_\text{biased } } \cdot \frac{ N_\text{events, biased} }{ N_\text{events, unbiased} }
\end{equation}
Poisson statistics are assumed for the number of events, i.e.\ number of different biasing indices for biased simulations (see also section \ref{sec_variance}), and these uncertainties are propagated to the efficiency boost.

Table~\ref{tab_eff_boost_gamma_neutrons} contains the achieved efficiency boosts in the validation studies for the reduced and the full shield thickness shown for gammas and neutrons in Fig.~\ref{fig_validation_gammas} and \ref{fig_validation_neutrons}, respectively. The efficiency boosts for multiple primaries per event turn out to be the same as for single primaries within statistical uncertainties and are not quoted in table~\ref{tab_eff_boost_gamma_neutrons}.

\begin{table}[t!]
\centering
\caption{Examples of the efficiency boost $\varepsilon$ achieved in validation studies.
The gamma simulations with four layers used a reduced lead shield thickness of 5~cm, while the 16 layers were run with the full lead shield thickness of 20~cm. Both had an importance layer thickness of 1.25~cm.
The neutron simulations were run with eight layers with the total shield thickness reduced to 60~cm and with 16 layers using the full thickness of 120~cm. Both had an importance layer thickness of 7.5~cm.
}
\label{tab_eff_boost_gamma_neutrons}
\begin{tabular}{ccr}
\toprule
Primary & Layers & \multicolumn{1}{c}{$\varepsilon$} \\
\midrule
$\gamma$ & 4 &  2.7751 $\pm$ 0.0073 \\
$\gamma$ & 16 & 1023 $\pm$ 25\hphantom{000.} \\
$n$ & 8  & 20.344 $\pm$ 0.021\hphantom{0} \\
$n$ & 16 & 478.8 $\pm$ 4.4\hphantom{000} \\
\bottomrule
\end{tabular}
\end{table}

In summary, in the idealized simulations throwing particles into the direction of the detectors without any holes (stems in Fig.~\ref{fig_supersim_supercdms}) in the shield, an efficiency boost of $\mathcal{O}(10^3)$ can be achieved for gammas and about 500 for neutrons.

\subsection{Optimal Settings}
\label{sec_optimal_settings}

For SuperCDMS' background simulations, various components are contaminated with different isotopes, decay chains and neutron spectra. Depending on the location, material and particle energy, customized importance biasing settings might be beneficial. For each of the requested background simulations, optimal settings for the importance layer thickness and number of importance layers are explored.

On one side, we need to achieve a sufficient efficiency boost to satisfy our statistics needs for background estimation and modeling. On the other side, if the settings are overstated, artificial peaks by Compton scattered gammas might show up in the validation spectra making it hard to verify if the biased simulations accurately reproduce the unbiased spectra. Even worse, if a very large number of particles needs to be tracked due to importance biasing, the simulation could exceed the allocated memory.

The optimal balance is reached when a particle has a survival probability of approximately 50\% when traversing one importance layer thickness in a given material. In that case, even with many importance layers, on average one particle will be recorded per event when the primary particle starts directly at the outermost importance layer and the surviving tracks are recorded at the innermost importance layer. In reality, gammas and neutrons are isotropically emitted in radioactive decays, i.e.\ they propagate through the importance layers in all directions and angles. Consequently, each particle travels a different distance to reach the next importance layer and has a different survival probability.

\begin{figure}[t!]
\centering
\includegraphics[width=\linewidth]{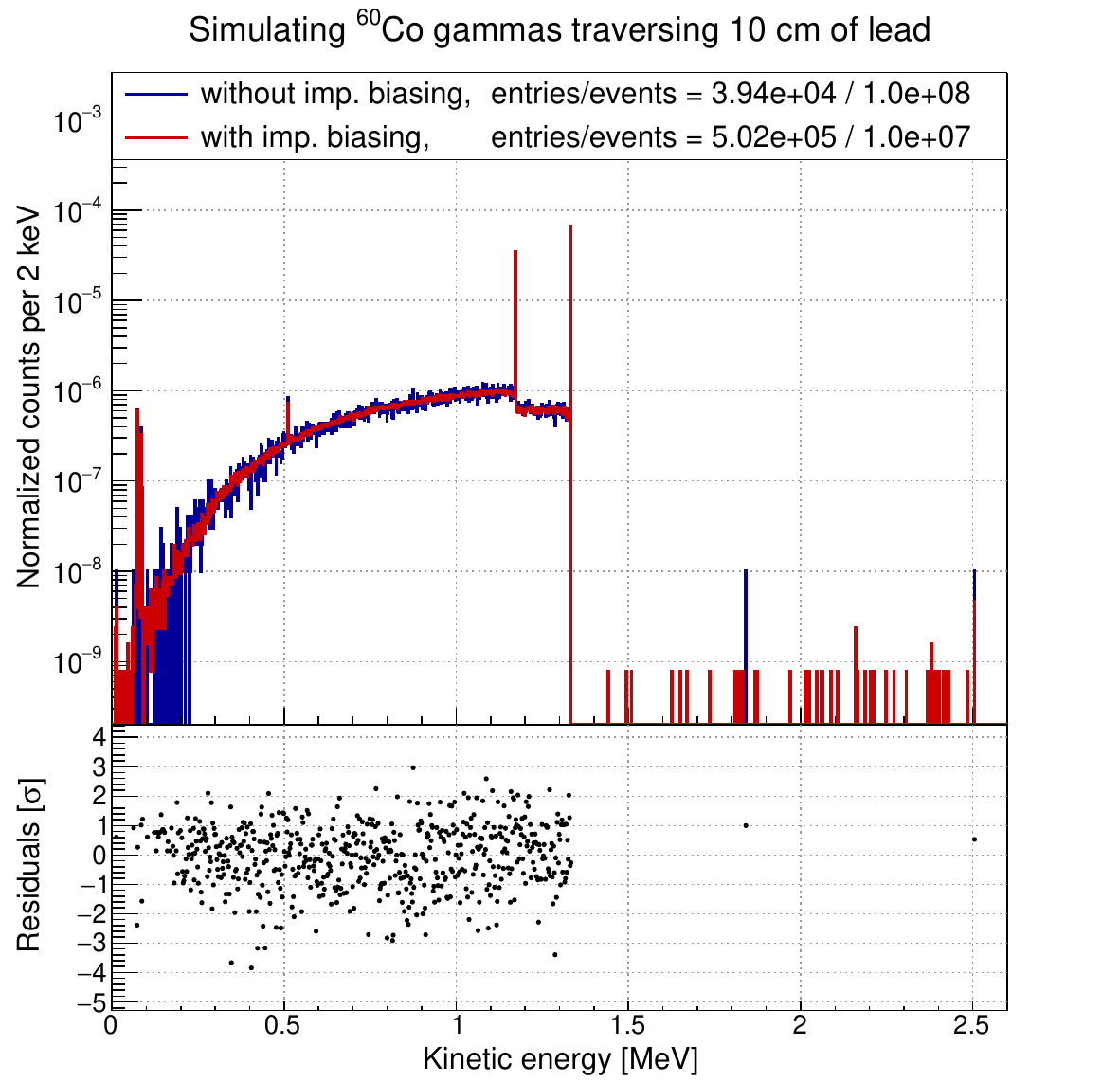}\\
\includegraphics[width=\linewidth]{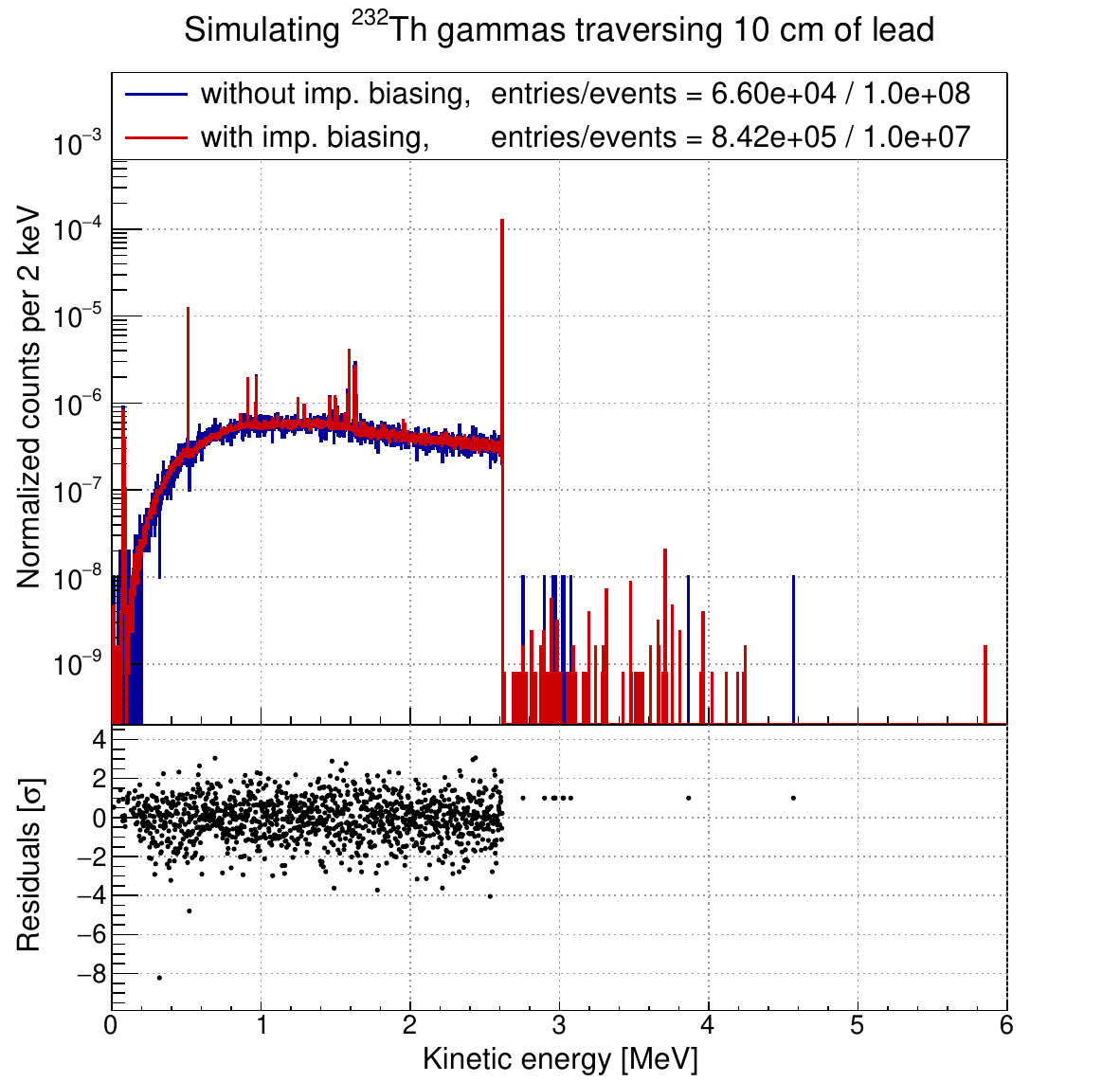}
\caption{
Simulated spectra generated with and without importance biasing for gammas emitted by \isotope[60]{Co} (top) and the \isotope[232]{Th} decay chain (bottom) in the radon barrier. In the case of the \isotope[232]{Th} decay chain, there is a small possibility that gammas from different isotopes have been recorded within the same event, leading to a larger total energy entry in the histogram.
The lead shield has been reduced to 10~cm for sufficient statistics in the unbiased spectra. Spectra for 20~cm of lead have also been checked to ensure that no artificial lines are introduced.
}
\label{fig_optimal_settings_gammas}
\end{figure}

For gammas, the attenuation length $\lambda$ or half-value layer $\text{HVL}=\ln (2) \cdot \lambda$ of a given energy in a material can be used as a first guess of an appropriate importance layer thickness. Due to the just described emission properties, the importance layer thickness should be a bit smaller than the HVL.
An additional complication is imposed by isotopes which emit multiple gammas of various energies and intensities.
For these cases, the best strategy seem to be to focus on the most prominent and high energy gammas emitted by an isotope or decay chain.
Fig.~\ref{fig_optimal_settings_gammas} shows example spectra for the optimal settings found for one isotope and one decay chain. The achieved efficiency boosts for several full scale background simulations are summarized in Table~\ref{tab_eff_boost_isotopes}.

\begin{table}[t!]
\centering
\caption{Efficiency boosts achieved in simulations contaminating the radon barrier (outer PE) far away from the stems with various isotopes or decay chains (neutrons). Tracks were propagated through 20~cm of lead (120~cm total shield thickness) and recorded at the inner PE shield (mu-metal). Simulations biasing gammas (neutrons) with 16 importance layers had an importance layer thickness of 1.25~cm (7.5~cm) and for 20 layers it was set to 1~cm. The latter was used for decay chains emitting only low energy gammas to improve the efficiency boost.
The \isotope[238]{U} decay is out of equilibrium in some of our materials, hence the upper chain simulation stops at \isotope[226]{Ra} and the lower chain is simulated separately.
The unbiased simulations for \isotope[235]{U} and \isotope[210]{Pb} were run for $10^9$ primary particles, but no track was recorded, hence a 90\% C.L. limit was set on the efficiency boost.
}
\label{tab_eff_boost_isotopes}
\begin{tabular}{ccr}
\toprule
Primary & Layers & \multicolumn{1}{c}{$\varepsilon$} \\
\midrule
\isotope[26]{Al}  & 16 & $(16.4 \pm 0.8)\cdot10^3$ \\
\isotope[60]{Co}  & 16 & $(49.8 \pm 7.6)\cdot10^3$  \\
\isotope[40]{K}   & 16 & $(22.7 \pm 6.3)\cdot10^3$  \\
\isotope[226]{Ra} & 16 & $(21.1 \pm 0.2)\cdot10^3$  \\
\isotope[232]{Th} & 16 & $(17.2 \pm 0.8)\cdot10^3$  \\
\isotope[235]{U}  & 16 & $> 17.9\cdot10^3$\\
\isotope[238]{U}  & 16 & $(28 \pm 14)\cdot10^3$  \\
\isotope[210]{Pb} & 20 & $> 2.8\cdot10^3$ \\
\midrule
$n\left(\isotope[232]{Th}\right)$ & 16 & 26.58 $\pm$ 0.06 \\
\bottomrule
\end{tabular}
\end{table}

\begin{figure}[t!]
\centering
\includegraphics[width=\linewidth]{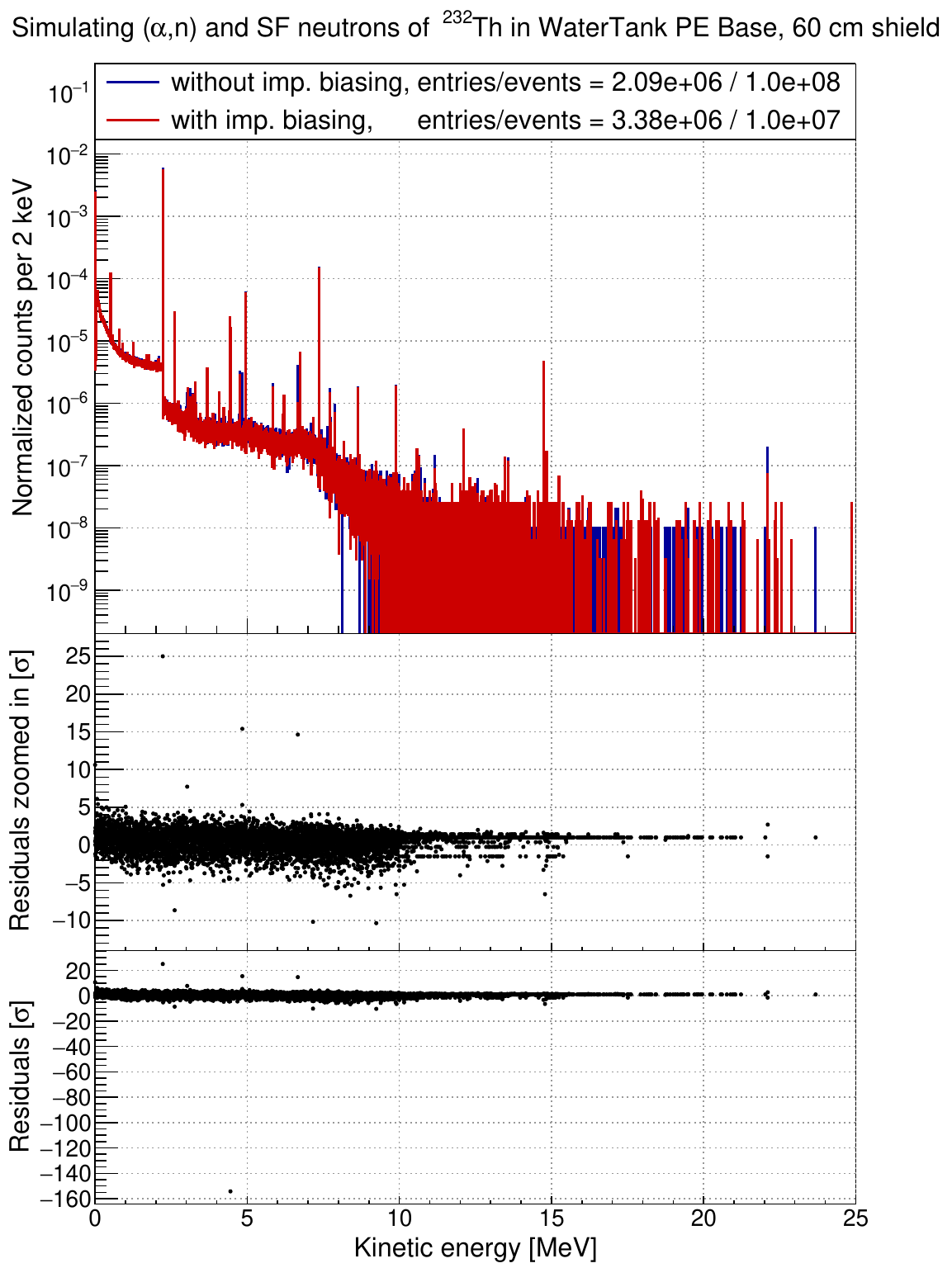}
\caption{
Simulated spectra for neutrons generated via the $(\alpha,n)$ process and spontaneous fission (SF) from \isotope[232]{Th} in the outer PE shield serving as a base of the water tank. The total shield thickness has been reduced to 60~cm. Spectra for 120~cm have been checked for consistency.
The same neutron physics as in Fig.~\ref{fig_validation_neutrons} apply here as well.
The residuals are very big for some specific lines, which can be explained. The most extreme outlier is at 4446.5~keV, which is a sum line of $\isotope[1]H(n,\gamma)\isotope[2]H$ (2223.25~keV) occurring twice within one event. Due to neutrons scattering much more often before this process, they likely got differing biasing indices and the correct topology cannot be reconstructed in the biased simulations (see section~\ref{sec_special_cases}).
As this one outlier dominates the axis scale of the residuals, a zoomed in version is additionally provided for the reader's convenience.
}
\label{fig_optimal_settings_neutrons}
\end{figure}

For neutrons, it is much more challenging to find optimal parameters, because their emission spectra is continuous and they penetrate through much more material than gammas. Consequently, the importance layers need to cover materials of different densities and atomic numbers. Depending on the shielding layers, it could even be beneficial to work with importance layers of various thicknesses which are adjusted to the different material's properties. For our purposes, we decided to keep a fixed importance layer thickness.
Fig.~\ref{fig_optimal_settings_neutrons} shows an example spectra for importance biasing settings that work well for neutron simulations. The efficiency boost achieved in a full shield simulation is included in Table~\ref{tab_eff_boost_isotopes}.

Note, that in this particular example of simulating neutrons from the \isotope[232]{Th} decay chain, the neutrons are started inside the SuperCDMS outer PE shield and 8 out of the 16 importance layers are overlaid with the outer PE shield. Hence, the efficiency boost is much lower because not all neutrons have to traverse through the full outer PE shield and with that only propagate through a lower number of importance layers.

\begin{table*}[t!]
\centering
\caption{
Efficiency boosts achieved in simulations run with and without importance biasing contaminating the SuperCDMS radon barrier with the primary isotopes.
Shown are the number of simulated events and the number of events making hits in at least one of the 24 SuperCDMS detectors.
For biased simulations, the number of detected different, unique biasing indices can be much larger than one for an event. The runtime impressively shows how the biased simulations need less CPU time while observing significantly more detector hits. 
}
\label{tab_eff_boost_det_hits}
\begin{tabular}{c|cc|cc|cc|r}
\toprule
\multirow{2}{*}{Primary} &
\multicolumn{2}{c|}{No. of sim. events} &
\multicolumn{2}{c|}{Events making hits} &
\multicolumn{2}{c|}{Runtime [h]} &
\multicolumn{1}{c}{\multirow{2}{*}{$\varepsilon~[10^3]$}} \\
 & imp. bias. & no bias & imp. bias. & no bias & imp. bias. & no bias & \\
\midrule
\multicolumn{8}{c}{Contaminating the radon barrier far away from the stems} \\
\midrule
\isotope[40]{K}   & $10^7$ & $2\cdot10^{10}$ & \hphantom{00}31   & \hphantom{0}1  & \hphantom{0}2.9  & \hphantom{0}3915 & $> 10.5$ \\
\isotope[60]{Co}  & $10^7$ & $10^{10}$       & \hphantom{0}134  & \hphantom{0}4  & 12.4 & \hphantom{0}5972 & $16.4 \pm 8.3$ \\
\isotope[232]{Th} & $10^7$ & $10^{10}$       & 1929 & 62 & 22.1 & 12056 &  $18.8 \pm 2.4$ \\
\midrule
\multicolumn{8}{c}{Contaminating the whole radon barrier including close to the stems} \\
\midrule
\isotope[40]{K}   & $10^7$ & $2\cdot10^{10}$ & \hphantom{00}35   & \hphantom{0}4  & \hphantom{0}2.6  & \hphantom{0}3454 & $12.4 \pm 6.5$ \\
\isotope[60]{Co}  & $10^7$ & $10^{10}$       & \hphantom{0}175  & 10 & 14.4 & \hphantom{0}5900 & $7.5 \pm 2.4$ \\
\isotope[232]{Th} & $10^7$ & $10^{10}$       & 2063 & 75 & 27.1 & 12123 & $ 13.7 \pm 1.6$ \\
\bottomrule
\end{tabular}
\end{table*}

If simulations are not run separately for neutrons, but instead neutrons would be generated by \textsc{Geant4} by spontaneous fission (SF) in radioactive decays or produced in $(\alpha,n)$ reactions, another level of complication would be introduced.
First, the number of produced neutrons is extremely small due to the low SF yield and the small $(\alpha,n)$ cross section in most materials. Therefore it is hard to collect sufficient statistics for energy deposits by neutrons and their secondary particles.
Second, in this case it would be better to bias gammas and neutrons in the same simulation to propagate both particle types through the shielding. But the settings would need to be adjusted for the particle type, i.e.\ different number and thicknesses of the importance layers, to achieve sensible efficiency boosts.

On the other hand, different importance layer settings in the same simulation would obstruct to combine cohesive event topologies of the different biased particle types.

\subsection{Application in background simulations}

For the final validation step, we are looking at the energy deposits in the sensitive detectors instead of analyzing the hits in the larger geometry volumes as discussed in the previous section. This will demonstrate the actual efficiency boost that we can achieve for some selected background simulations. Due to the limited statistics in the unbiased simulations, only a few isotopes have been selected for this study and no spectra are shown. 

The SuperCDMS geometry is constructed with the full shield thickness and the importance layers are arranged in the same way as in the actual background simulations.
In the previous simulation studies, primary particles were always sampled far away from the stems to avoid statistical artifacts.  
For comparison, Table~\ref{tab_eff_boost_det_hits} shows the results of the simulations sampling primaries in the SuperCDMS radon barrier for both scenarios, avoiding and including the stems (see Fig.~\ref{fig_supersim_supercdms}).

Particles generated close to the stems have a higher probability of reaching the detectors, thus the statistics for contaminating the whole SuperCDMS radon barrier are slightly better, but the statistical uncertainties are still large due to the small amount of detector hits observed in the unbiased simulations. For the case of only one event making hits in a detector, a 90\% C.L. limit has been calculated using the Feldman-Cousins method for Poisson signals \cite{Feldman-Cousins}.

Looking at the number of events making hits in the detectors, Table~\ref{tab_eff_boost_det_hits} shows very clearly why all the validations discussed above were only feasible with a reduced shield thickness and recording tracks close to the innermost importance layer. The efficiency boosts give an indication what can be achieved in background simulation productions where the stems cannot be avoided. Nevertheless, the numbers for the efficiency boosts are on the same order of magnitude as shown in Table~\ref{tab_eff_boost_isotopes} avoiding the stems.

\section{Variance observations}
\label{sec_variance}

During the simulation studies where tracks are recorded near the innermost importance layer (i.e.\ inner PE shield for gammas), it was noticed that while the energy spectra generated with and without importance biasing match very well  with each other, the variance in the biased spectra seems to be higher than expected. To investigate this quantitatively, the distribution of the number of different biasing indices recorded for gammas at the inner PE shield was plotted for biased simulations, and the distribution of the number of recorded events was plotted for unbiased simulations.

\begin{figure}[t!]
\centering
\includegraphics[width=\linewidth]{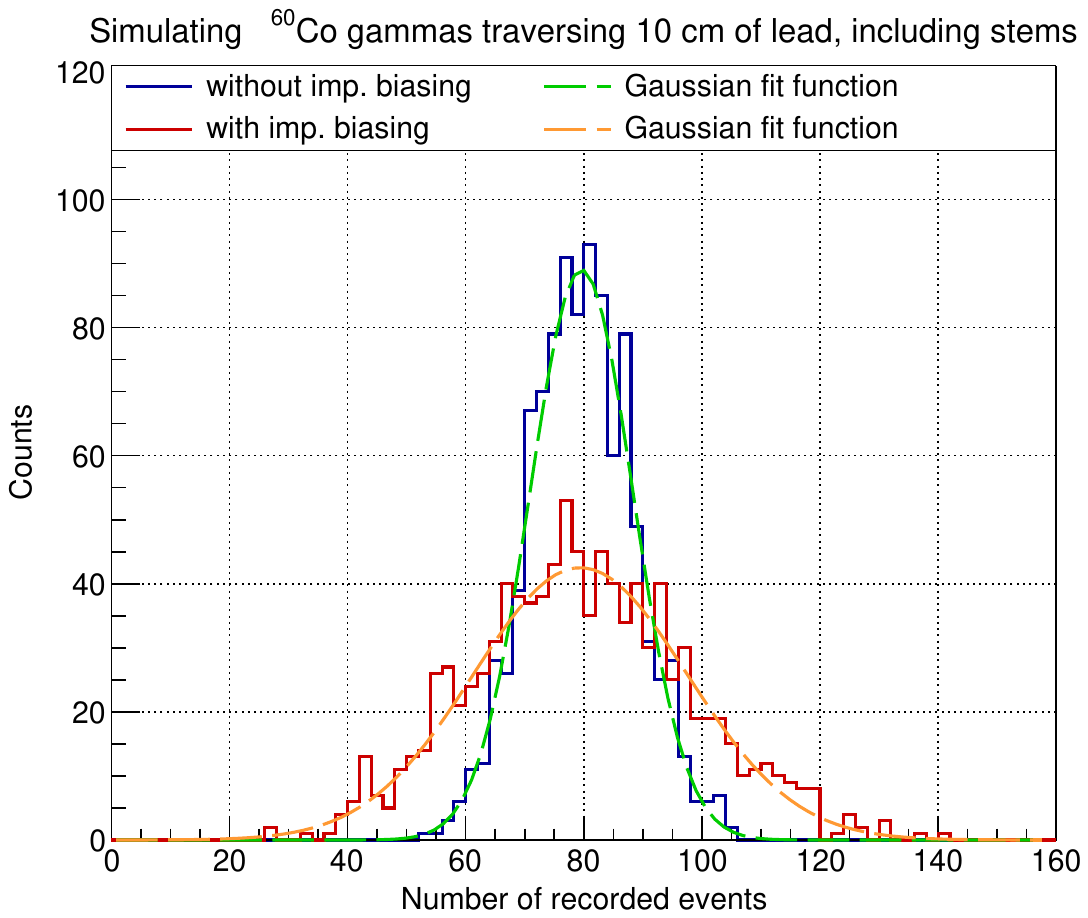}\\
\includegraphics[width=\linewidth]{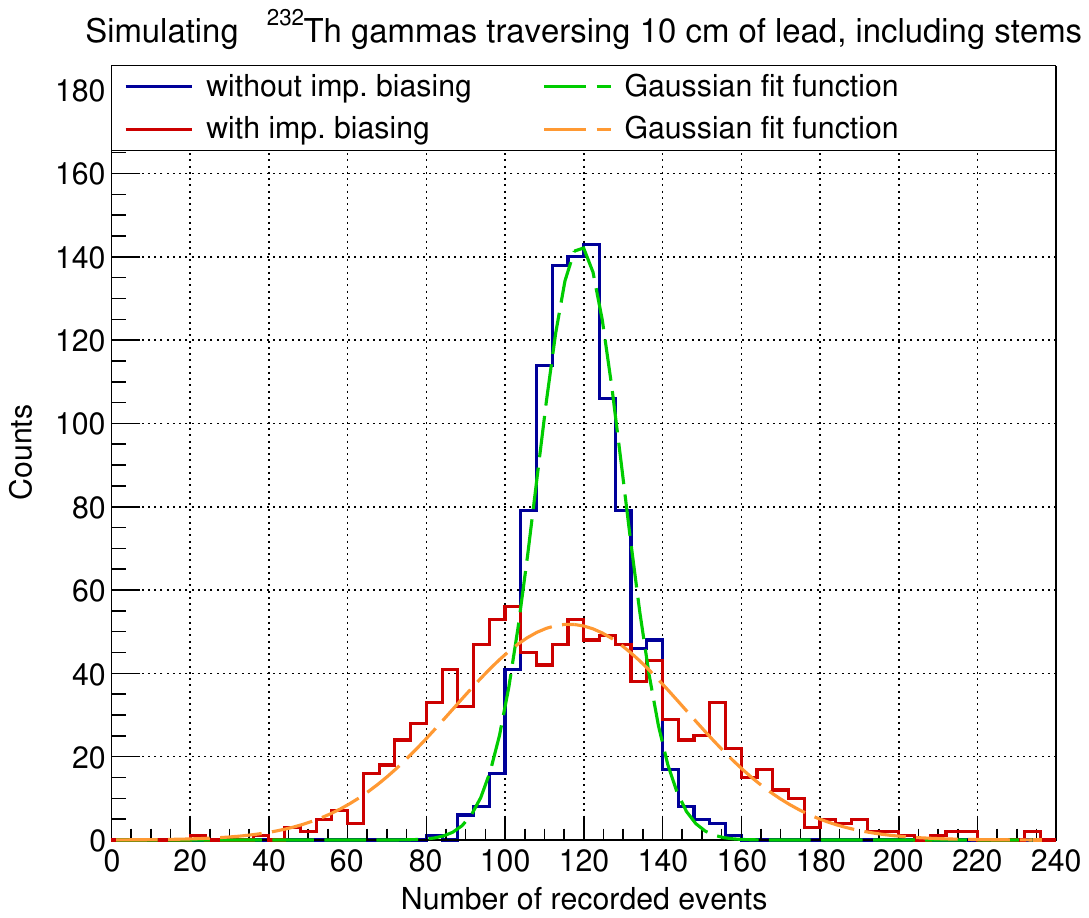}
\caption{
Simulations run with and without importance biasing contaminating the radon barrier (including close to the stems) with \isotope[60]{Co} (top) and \isotope[232]{Th} (bottom) and recording tracks at the inner PE shield.
The lead shield has been reduced to 10~cm for sufficient statistics in the unbiased spectra. The biased simulations were run with eight importance layers spanning the lead shield with an importance layer thickness of 1.25~cm.
With eight importance layers, there are $2^{L-1} = 128$ different track topologies.
The number of recorded different biasing indices per $10^3$ simulated events have been plotted for the biased simulations and the number of recorded events per $128\cdot10^3$ simulated events have been plotted for the unbiased simulations. Each histogram contains $10^3$ entries.
The distributions have been fit with a Gaussian function. See Table~\ref{tab_variance} for fit results.
}
\label{fig_variance_8layers}
\end{figure}

In Fig.~\ref{fig_variance_8layers}, one can see that the biased simulations have a larger variance compared to the unbiased version.  
We do see the exact same effect when contaminating the SuperCDMS radon barrier far away from the stems with the respective isotopes, hence, the stem holes are not responsible for this effect.
Also, the effect seems to be stronger for \isotope[232]{Th} than for \isotope[60]{Co}.
One hypothesis is that this is related to the higher energies of the gammas emitted by the \isotope[232]{Th} decay chain (2.61~MeV from \isotope[208]{Tl}) compared to \isotope[60]{Co} (1.17~MeV, 1.33~MeV).

\begin{figure}[t!]
\centering
\includegraphics[width=\linewidth]{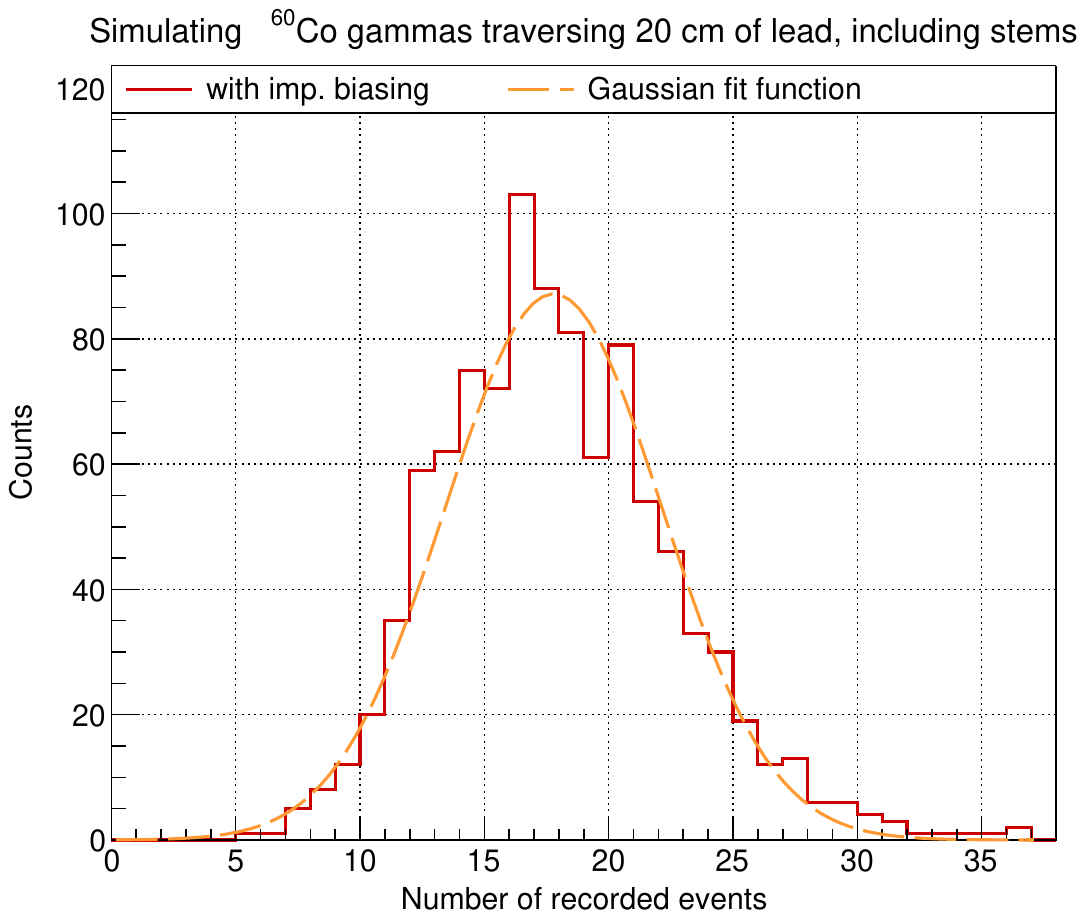}\\
\includegraphics[width=\linewidth]{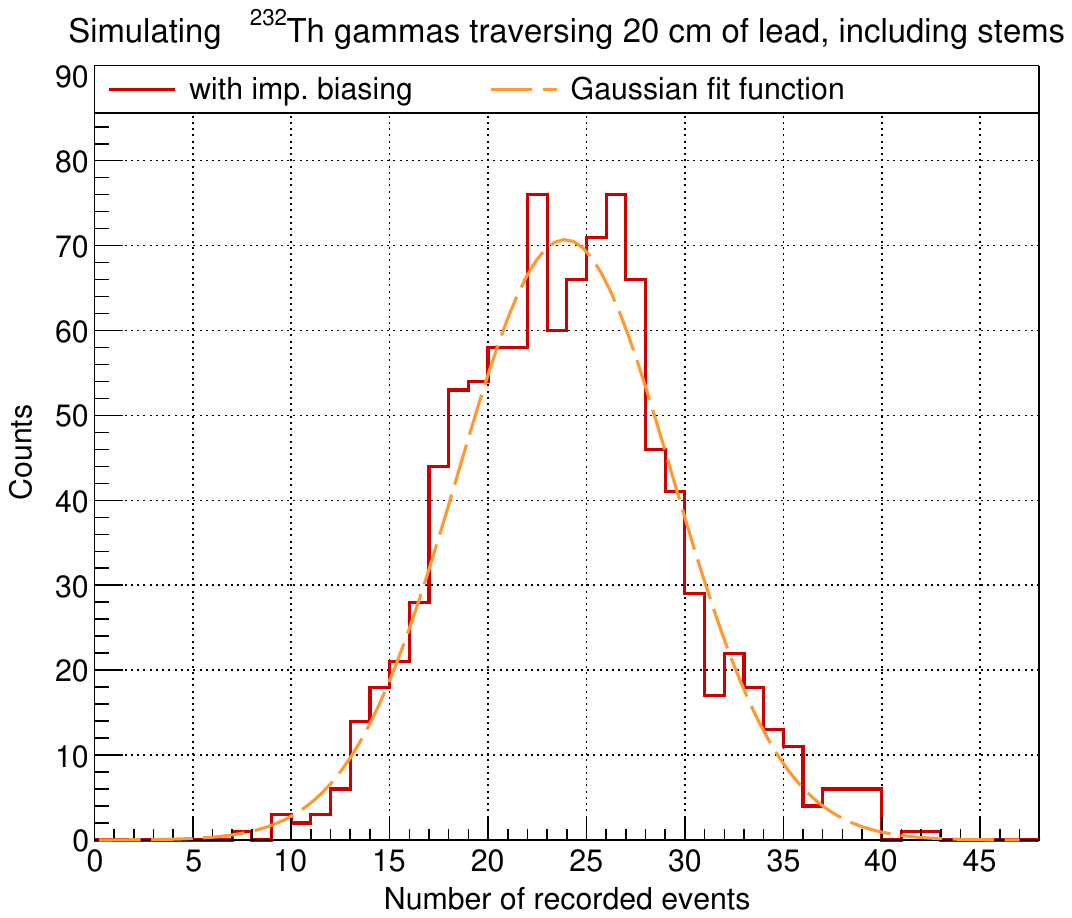}
\caption{
Simulations runs with importance biasing contaminating the radon barrier (including close to the stems) with \isotope[60]{Co} (top) and \isotope[232]{Th} (bottom) and recording energy depositions in the 24 SuperCDMS detectors. The lead shield has its full thickness of 20~cm. The simulations were run with 16 importance layers with a thickness of 1.25~cm each.
The number of recorded different biasing indices per $10^6$ (\isotope[60]{Co}) and $10^5$ (\isotope[232]{Th}) simulated events have been plotted. Each histogram contains $10^3$ entries.
The simulated distributions have been fit with a Gaussian function. See Table~\ref{tab_variance} for fit results.
}
\label{fig_variance_16layers_dethits}
\end{figure}

In order to investigate the variance in real applications such as background simulations, we ran simulations with importance biasing in the same way as before, with the same reduced shield of 10~cm and only eight importance layers (Fig.~\ref{fig_variance_8layers}) and also with the full shield of 20~cm and 16 importance layers, but recording the detector hits (Fig.~\ref{fig_variance_16layers_dethits}). Due to the limited statistics no unbiased simulation is available for comparison for the latter case.

The properties of the simulated distributions (Fig.~\ref{fig_variance_8layers} and Fig.~\ref{fig_variance_16layers_dethits}) and the parameters of Gaussian fits to the simulation results are listed in Table~\ref{tab_variance}.

While the mean number of different biasing indices $\mathcal{B}$ recorded at the inner PE shield agrees very well with the recorded number of events in the unbiased simulations, the standard deviation is roughly twice (\isotope[60]{Co}) or three times (\isotope[232]{Th}) larger in the biased simulations (rows 1--4 in table~\ref{tab_variance}). The Gaussian fit parameters show the same relations.
For a distribution which perfectly follows Poisson statistics with the standard deviation $\sigma$ and the mean $\mu$, the ratio of $\sigma / \sqrt{\mu} $ is expected to be one. This condition is well satisfied by the unbiased simulations. The biased simulations show larger ratios due to the larger standard deviation of the distributions. For all four simulations the goodness of fit described by $\chi^2 / \text{ndf}$ and the associated p-value show that the Gaussian fit is a good approximation of the respective distribution.

\begin{table*}[t!]
\centering
\caption{
Comparison of the mean $\mu$ and standard deviation $\sigma$ of the simulated distribution of the number of different biasing indices with a Gaussian fit of the distribution.
The ratio of $\sigma / \sqrt{\mu} $ using the Gaussian fit parameters show how well the simulated distribution follows Poisson statistics.
Unbiased simulations (layers $-$) and biased simulations are the ones described in Fig.~\ref{fig_variance_8layers} and \ref{fig_variance_16layers_dethits}, respectively, i.e.\ in the first four rows are tracks recorded at the inner PE shield and in the last four rows are energy depositions in the detectors.
}
\label{tab_variance}
\begin{tabular}{cc|cc|cc|ccc}
\toprule
\multirow{2}{*}{Primary} &
\multirow{2}{*}{Layers} &
\multicolumn{2}{|c|}{Num. of diff. $\mathcal{B}$ making hits} &
\multicolumn{2}{|c|}{\centering{Gaussian fit parameters}} &
\multirow{2}{*}{$ \dfrac{\sigma}{\sqrt{\mu}}$} &
\multirow{2}{*}{$\dfrac{\chi^2}{\text{ndf}}$} &
\multirow{2}{*}{p-value} \\
 & & $\mu$ & $\sigma$ & $\mu$ & $\sigma$\\
\midrule
\multicolumn{9}{c}{Tracks recorded at inner PE shield} \\
\midrule
\isotope[60]{Co} & $-$ & \phantom{0}79.21 $\pm$ 0.28 & \phantom{0}8.81 $\pm$ 0.20 & \phantom{0}79.61 $\pm$ 0.29 & \phantom{0}8.77 $\pm$ 0.22 & 0.983 & 0.971 & 0.502 \\
\isotope[60]{Co} & 8   & \phantom{0}79.33 $\pm$ 0.61 & 19.39 $\pm$ 0.43 & \phantom{0}79.40 $\pm$ 0.62 & 18.16 $\pm$ 0.49 & 2.038 & 0.869 & 0.723 \\ 
\isotope[232]{Th} & $-$ & 118.77 $\pm$ 0.35 & 11.22 $\pm$ 0.25 & 119.04 $\pm$ 0.36 & 10.97 $\pm$ 0.29 & 1.006 & 1.331 & 0.167 \\ 
\isotope[232]{Th} & 8   & 116.55 $\pm$ 0.99 & 31.21 $\pm$ 0.70 & 116.23 $\pm$ 1.00 & 29.49 $\pm$ 0.74 & 2.735 & 1.109 & 0.288 \\ 
\midrule
\multicolumn{9}{c}{Energy depositions recorded in detectors} \\
\midrule
\isotope[60]{Co}  & 8  & \phantom{0}38.56 $\pm$ 0.20 & \phantom{0}6.39 $\pm$ 0.14 & 39.07 $\pm$ 0.21 & \phantom{0}6.29 $\pm$ 0.15 & 1.006 & 1.059 & 0.375 \\
\isotope[60]{Co}  & 16  & \phantom{0}18.64 $\pm$ 0.61 & 19.15 $\pm$ 0.43 & 17.79 $\pm$ 0.15 & \phantom{0}4.37 $\pm$ 0.11 & 1.037 & 1.391 & 0.079 \\ 
\isotope[232]{Th} & 8  & \phantom{0}11.21 $\pm$ 0.11 & \phantom{0}3.43 $\pm$ 0.08 & 11.66 $\pm$ 0.11 & \phantom{0}3.41 $\pm$ 0.08 & 0.998 & 0.765 & 0.752 \\
\isotope[232]{Th} & 16  & \phantom{0}23.62 $\pm$ 0.19 & \phantom{0}5.86 $\pm$ 0.13 & 23.90 $\pm$ 0.18 & \phantom{0}5.47 $\pm$ 0.14 & 1.118 & 1.074 & 0.356 \\
\bottomrule
\end{tabular}
\end{table*}

Table~\ref{tab_variance} also shows that the mean number of different biasing indices $\mathcal{B}$ recorded in the detectors agrees with the respective Gaussian fit mean. However, for \isotope[60]{Co} the standard deviation of the simulated distribution is very large (sixth row in table~\ref{tab_variance}), which is not obvious in Fig.~\ref{fig_variance_16layers_dethits}. This is caused by five events out of $10^9$ simulated events, for which in total 1269 different $\mathcal{B}$ have been recorded. The primary isotopes for all five of these events have been started closer than 50~cm to the C-Stem hole, giving them a geometrical advantage to reach the detectors.
These outliers are specific to SuperCDMS' geometry and with sufficient statistics in an unbiased simulation these should show up as well. However, considering the 16 importance layers in these simulations, i.e. $2^{L-1} = 32768 $ different track topologies, one would expect to see about ten such outliers in $32 \cdot 10^{10} $ unbiased events, which would take on the order of 10 CPU years. In any case, we do not consider these outliers problematic due to their geometrical nature.

The Gaussian fit parameters of the biased simulations with 8 and 16 importance layers, where the mean number of different biasing indices $\mathcal{B}$ have been recorded in the detectors, show that the plotted distributions without outliers follow Poisson statistics quite well (rows 5--8 in table~\ref{tab_variance}). The effect of the larger variance is only visible when recording tracks close to the innermost importance layer, while these features are washed out when looking at the detector hits.

With importance biasing, the number of primaries to achieve the same number of detector hits compared to an unbiased simulation is significantly smaller, because the number of particle tracks per primary is much larger in a biased simulation.
Consequently, the initial variance of the primary particles (starting position, emission direction and energy in a decay) is relatively larger compared to the number of particles reaching the innermost importance layer. This peculiarity of importance biasing becomes more apparent when the settings (layer thickness, number of layers) are chosen to achieve a higher efficiency and it is less obvious when the settings are rather moderate.
In fact, the ideal is that every primary contributes exactly one event that reaches the innermost importance layer, i.e.\ the mean number $\overline{N_\mathcal{B}}$ of different biasing indices $\mathcal{B}$ per primary is one.
However, perfect settings are impossible when different gamma energies need to be covered at the same time as pointed out in section~\ref{sec_optimal_settings}.
In any case, if the rim of the innermost importance layer is far enough away from the sensitive detectors, particles can be absorbed on their way to the detectors and it is less likely that particles with different biasing indices, which are caused by the same primary, hit the detector.

As a consequence, the importance layers, in particular the innermost importance layer, should not be located in proximity to the sensitive detectors to avoid non-Poisson distributed hits.
This is important to know when the uncertainty of the calculated rate or the simulated energy spectra are relevant for constructing a background model that relies on the uncertainty of each energy bin.

\section{Conclusion}

\textsc{Geant4} provides various techniques to significantly improve the efficiency of simulations. Among the available techniques is importance biasing which has been investigated for the application in background simulations for SuperCDMS.
In this work, importance biasing has been used in simulations of radioactive backgrounds intrinsic in the experiment's materials, but the application of importance biasing can be extended to neutrons and gammas emitted by the SNOLAB cavern as well as cosmic muons.

Dedicated simulations have been performed to explore optimal settings for all isotopes and decay chains relevant to SuperCDMS. Additionally, the application of importance biasing to neutrons has been investigated and proven feasible.

With the determined optimal settings for importance biasing, future background simulations for SuperCDMS will consume orders of magnitude less computing time, while at the same time achieving the statistics requirements for the detected energy spectra to develop a proper background model.

A larger variance in importance biasing simulations has been observed when recording tracks close to the innermost importance layer. Fortunately, this effect is not visible in detector spectra and distributions simulated with importance biasing are following Poisson statistics when observing the detector hits directly.

On a final note, \textsc{Geant4}'s importance biasing enables us to reduce the carbon emission of our simulations by needing less computing time to meet our objectives.

\section{Acknowledgments}

The authors would like to thank the SuperCDMS collaboration for their continuous support and friendly work environment.

This research was enabled in part by support provided by Compute Ontario (\href{https://www.computeontario.ca/}{computeontario.ca}) and the Digital Research Alliance of Canada (\href{https://alliancecan.ca/en}{alliancecan.ca}). 
The simulations performed in this work have utilized allocations on computing clusters provided by the Digital Research Alliance of Canada.

Funding and support were received from NSERC Canada, the Canada First Excellence Research  Fund, the Arthur B. McDonald Canadian Astroparticle Physics Research Institute, the Canada Foundation for Innovation, the National Science Foundation, the U.S. Department of Energy (DOE).

This material is based in part upon work supported by the National Science Foundation under Grant No.\ 1707704. Any opinions, findings, and conclusions or recommendations  expressed in this material are those of the authors and do not necessarily reflect the views of the National Science Foundation.


\begin{thebibliography}{00}

\hbadness=10000

\bibitem{Geant4-2016} J. Allison et al.,
Recent Developments in \textsc{Geant4},
Nucl. Instrum. Meth. A 835 (2016) 186-225.
DOI: \href{https://doi.org/10.1016/j.nima.2016.06.125}
{\nolinkurl{10.1016/j.nima.2016.06.125}}.

\bibitem{Geant4-2006} J. Allison et al.,
\textsc{Geant4} Developments and Applications,
IEEE Trans. Nucl. Sci. 53 (2006) 270-278.
DOI: \href{https://doi.org/10.1109/TNS.2006.869826}
{\nolinkurl{10.1109/TNS.2006.869826}}.

\bibitem{Geant4-2003} S. Agostinelli et al.,
\textsc{Geant4} - A Simulation Toolkit,
Nucl. Instrum. Meth. A 506 (2003) 250-303.
DOI: \href{https://doi.org/10.1016/S0168-9002(03)01368-8}
{10.1016/S0168-9002(03)01368-8}.

\bibitem{SNO-2009} B. Aharmim et al. (SNO Collaboration),
Measurement of the cosmic ray and neutrino-induced muon flux at the Sudbury neutrino observatory,
Phys. Rev. D 80 (2009) 012001.
DOI: \href{https://doi.org/10.1103/PhysRevD.80.012001}
{10.1103/PhysRevD.80.012001}.

\bibitem{MCNP-Man} MCNP\textsuperscript{\textregistered} Code Version 6.3.0 Theory \& User Manual,
DOI: \href{https://doi.org/10.2172/1889957}
{10.2172/1889957},
accessed 15 Apr. 2025.

\bibitem{FLUKA} Overview of the FLUKA code,
Annals of Nuclear Energy 82, 10-18 (2015),
DOI: \href{https://doi.org/10.1016/j.anucene.2014.11.007}
{10.1016/j.anucene.2014.11.007},
URL: \href{https://fluka.cern}{\nolinkurl{https://fluka.cern}},
accessed 15 Apr. 2025.

\bibitem{Geant4-AppDev} \textsc{Geant4}'s Book For Application Developers,
URL: \href{https://geant4-userdoc.web.cern.ch/UsersGuides/ForApplicationDeveloper/html/index.html}
{\nolinkurl{https://geant4-userdoc.web.cern.ch/UsersGuides/ForApplicationDeveloper/html/index.html}},
accessed 29 Dec. 2024.

\bibitem{TAUP-Proc} B.~Zatschler on behalf of the SuperCDMS Collaboration, Background simulations for the SuperCDMS experiment – Efficient \textsc{Geant4} simulations using Importance Biasing, PoS TAUP2023 (2024) 024.
DOI: \href{https://doi.org/10.22323/1.441.0024 }{10.22323/1.441.0024}.

\bibitem{VIEWS-Talk} B.~Zatschler,
Application of \textsc{Geant4}'s Importance Biasing in radiogenic background simulations, VIEWS24,
\href{https://indico.cern.ch/event/1275551/contributions/5858758/}
{\nolinkurl{https://indico.cern.ch/event/1275551/contributions/5858758/}},
accessed 28 Nov. 2024.

\bibitem{David-Thesis} Pedreros, D. S.,
Position measurement of the SuperCDMS HVeV detector and implementation of an importance sampling algorithm in the SuperCDMS simulation software,
Master's thesis (2023).
URL: \href{https://hdl.handle.net/1866/32096}
{\nolinkurl{https://hdl.handle.net/1866/32096}}.

\bibitem{Geant4-Phys} \textsc{Geant4}'s Physics List Guide,
URL: \href{https://geant4-userdoc.web.cern.ch/UsersGuides/PhysicsListGuide/html/index.html}
{\nolinkurl{hhttps://geant4-userdoc.web.cern.ch/UsersGuides/PhysicsListGuide/html/index.html}},
accessed 29 Dec. 2024.

\bibitem{SOURCES4C} W.B.~Wilson et al.,
SOURCES 4A: A Code for Calculating ($\alpha$,n), Spontaneous Fission, and Delayed Neutron Sources and Spectra,
LA-13639 (1999)
DOI: \href{https://doi.org/10.2172/15215}
{10.2172/15215}.

\bibitem{Feldman-Cousins} Feldman, G. J. and Cousins, R. D.,
Unified approach to the classical statistical analysis of small signals,
Phys. Rev. D 57 (1998) 3873.
DOI: \href{https://doi.org/10.1103/PhysRevD.57.3873}
{\nolinkurl{10.1103/PhysRevD.57.3873}}.

\end{thebibliography}
\end{document}